\documentclass[sigconf]{acmart}

\usepackage[english]{babel}
\usepackage{blindtext}
\usepackage[ruled, vlined, linesnumbered]{algorithm2e}
\SetKwComment{Comment}{// }{}
\SetKwInOut{Input}{Input}
\SetKwInOut{Output}{Output}
\renewcommand\footnotetextcopyrightpermission[1]{} % removes footnote with conference info
\setcopyright{none}
\acmDOI{}

\acmISBN{}

\acmPrice{}

\usepackage{hyperref}

\usepackage{color}
\usepackage{tikz}
\usepackage[tight]{subfigure}
\usepackage{comment}
\usepackage{amsmath}
\usepackage{enumerate}
\usepackage{multirow}
\usepackage{pifont}
\usepackage{hhline}

\usepackage{amssymb}

\usepackage{enumerate}
\usepackage{enumitem}

\usepackage{xurl}
\usepackage{slashbox}
\usepackage[table]{xcolor}
\usepackage{amsmath}
\usepackage{xspace}
\usepackage[font=small]{caption}
\usepackage[noend]{algpseudocode}
\usepackage[capitalize]{cleveref}
\makeatletter
\def\BState{\State\hskip-\ALG@thistlm}
\makeatother

\algdef{SE}[SUBALG]{Indent}{EndIndent}{}{\algorithmicend\ }%
\algtext*{Indent}
\algtext*{EndIndent}
\algrenewcommand\algorithmicindent{1.0em}%

\usepackage[normalem]{ulem}
\usepackage{lipsum}

\newif\ifcomm
\ifcomm
\else
\commfalse
\fi
\ifcomm
	\newcommand{\mycomm}[3]{{\footnotesize{{\color{#2} \textbf{[#1: #3]}}}}}
	\newcommand{\CRdel}[1]{\textcolor{red}{\sout{#1}}}

        \newcommand{\remove}[1]{\textcolor{red}{\sout{#1}}}
\else
    \newcommand{\mycomm}[3]{}
    \newcommand{\CRdel}[1]{}

    \newcommand{\remove}[1]{}
\fi

\definecolor{Wenchen}{RGB}{200,0,200}
\definecolor{gingfung}{RGB}{255,125,0}

\newcommand{\Wenchen}[1]{\mycomm{Wenchen}{Wenchen}{#1}}
\newcommand{\gingfung}[1]{\mycomm{gingfung}{gingfung}{#1}}

\hypersetup{pdfstartview=FitH,pdfpagelayout=SinglePage}

\newcommand{\subp}{\noindent\textbf}

\newcommand{\ie}{\textit{i.e.}}
\newcommand{\eg}{\textit{e.g.}}

\newcommand{\sysname}{Lynx\xspace}

\begin{abstract}
    Long-context inference is increasingly common in large language model (LLM) serving, driven by retrieval-augmented generation and agentic systems. In disaggregated inference, these workloads require transferring large Key-Value (KV) caches across the network, where decoding cannot begin until the transfer completes. Recent KV quantization techniques reduce data volume and alleviate this bottleneck, but existing schemes fail to achieve both low network-exposed latency and high inference accuracy.

    We challenge the assumption that the KV cache is an indivisible unit that must be fully received before use. We leverage the observation that different bits in the KV cache contribute unequally to attention computation and inference precision: the most significant bits capture the coarse structure of attention and the least significant bits refine precision. This property enables partial use of the KV cache during decoding. We present Lynx, a system that enables progressive, split-stream KV transfer by partitioning the KV cache into a high-priority Anchor stream carrying the most significant bits and a low-priority Residual stream carrying remaining precision. Decoding begins upon receipt of the Anchor stream and proceeds speculatively while the Residual stream is transferred concurrently, followed by verification that ensures equivalence to higher-precision decoding. 

    Across multiple models and serving workloads, Lynx achieves Time-to-First-Token (TTFT) comparable to aggressive 4-bit KV quantization, while matching the accuracy of high-precision (BF16) inference, improving TTFT over standard 8-bit KV quantization by up to $1.43\times$ and improving accuracy over state-of-the-art by up to $5.1\%$.

\end{abstract}

\begin{document}

% \conferenceinfo{HotNets 2021} {}
% \CopyrightYear{2021}
% \crdata{X}
% \date{}

%%%%%%%%%%%% THIS IS WHERE WE PUT IN THE TITLE AND AUTHORS %%%%%%%%%%%%

\title[Lynx: Progressive Speculative Quantization]{Lynx: Progressive Speculative Quantization for accelerating KV Transfer in Long-Context Inference}
% //FRANCIS: Enabling distributed message passing at switches}

% \newif\ifauthor
% \authortrue
% \ifauthor
% \else
% \authorfalse
% \fi

% %Usenix format
% \ifauthor

\author{Wenchen Han}
\affiliation{%
  \institution{University College London}%
  % \city{London}
  \country{}%  
}

\author{Gingfung Matthew Yeung} 
\affiliation{%
  \institution{Huawei}%
  \country{}%  
}

\author{Marco Barletta}
\affiliation{%
  \institution{Huawei}%
  \country{}%  
}

%\authornote{Michael Mitzenmacher was supported in part by NSF grants CCF-2101140, CNS-2107078, and DMS-2023528.
%}

\author{William Toner}
\affiliation{%
  \institution{Huawei}%
  % \city{London}
  \country{}%  
}  

\author{Amory Hoste}
\affiliation{%
  \institution{Huawei}%
  % \city{London}
  \country{}%  
} 

\author{Adam Barker}
\affiliation{%
  \institution{Huawei}%
  % \city{London}
  \country{}%  
} 
% \else{
% \author{Paper \#141, 12 pages body, 15 pages total}
% }

\renewcommand{\shortauthors}{W. Han, G. Yeung, M. Barletta, W. Toner, A. Hoste, A. Barker}

\maketitle

\section{Introduction}
%\marco{IMHO: main drawbacks intro atm. See side comments} %MARCO: atm is that 1) is very wordy. it takes more than a page to understand we do spec stuff to verify the draft tokens generated with the first stream. From the end of first column on  could be more direct. 2) a little generalization: what are the general implications of the paper? Could be something like "We introduce a new class of algorithms that use speculative stuff with verification at system level (i.e., system-configurations of the same AI model - quantization) instead of AI level (i.e., the usual specDec)"

%% ADAM EDITS

%how to mention 4b ??
%\Wenchen{Unify the terms of MSB-4b, LSB-4b.}

Large Language Model (LLMs) inference is increasingly dominated by long-context workloads, where prompts of tens to hundreds of thousands of tokens are common in retrival-augmented generation~\cite{yao2025cacheblend}, agentic systems~\cite{claudecode}, and code intelligence~\cite{yang2024swe}. State-of-the-art models such as Gemini 3.0~\cite{gemini3}, Qwen 3~\cite{qwen3}, and DeepSeek 3.2~\cite{liu2025deepseek} already support context windows of more than 1 million tokens.

In order to sustain throughput under these workloads, modern serving systems disaggregate inference into a compute-bound prefill stage and a memory-bound decode stage, placing them on separate accelerator instances~\cite{nvidia2025dynamo,liu2024deepseek,zuo2025serving}. While prefill-decode disaggregation improves hardware utilization, it introduces a fundamental bottleneck: the transfer of Key-Value (KV) cache\footnote{KV cache encodes the context information of historical tokens and is used for the \textit{attention computation} in LLM's forward pass computation.} between instances~\cite{yue2025rtt}. 

The KV cache grows linearly with context length and model depth, reaching tens of gigabytes for a single long-context request~\cite{qin2025mooncake,yang2025beluga}. Even on high-bandwidth interconnects, transferring this state introduces substantial latency that directly inflates Time-to-First-Token (TTFT) and delays decoding. As context lengths continue to scale, KV transfer latency, not compute, becomes the dominant limiter of end-to-end inference performance in disaggregated deployments~\cite{zhang2025hack,zhu2025megascale}. 

\begin{figure}[t!]
    \centering
    \includegraphics[width=\linewidth]{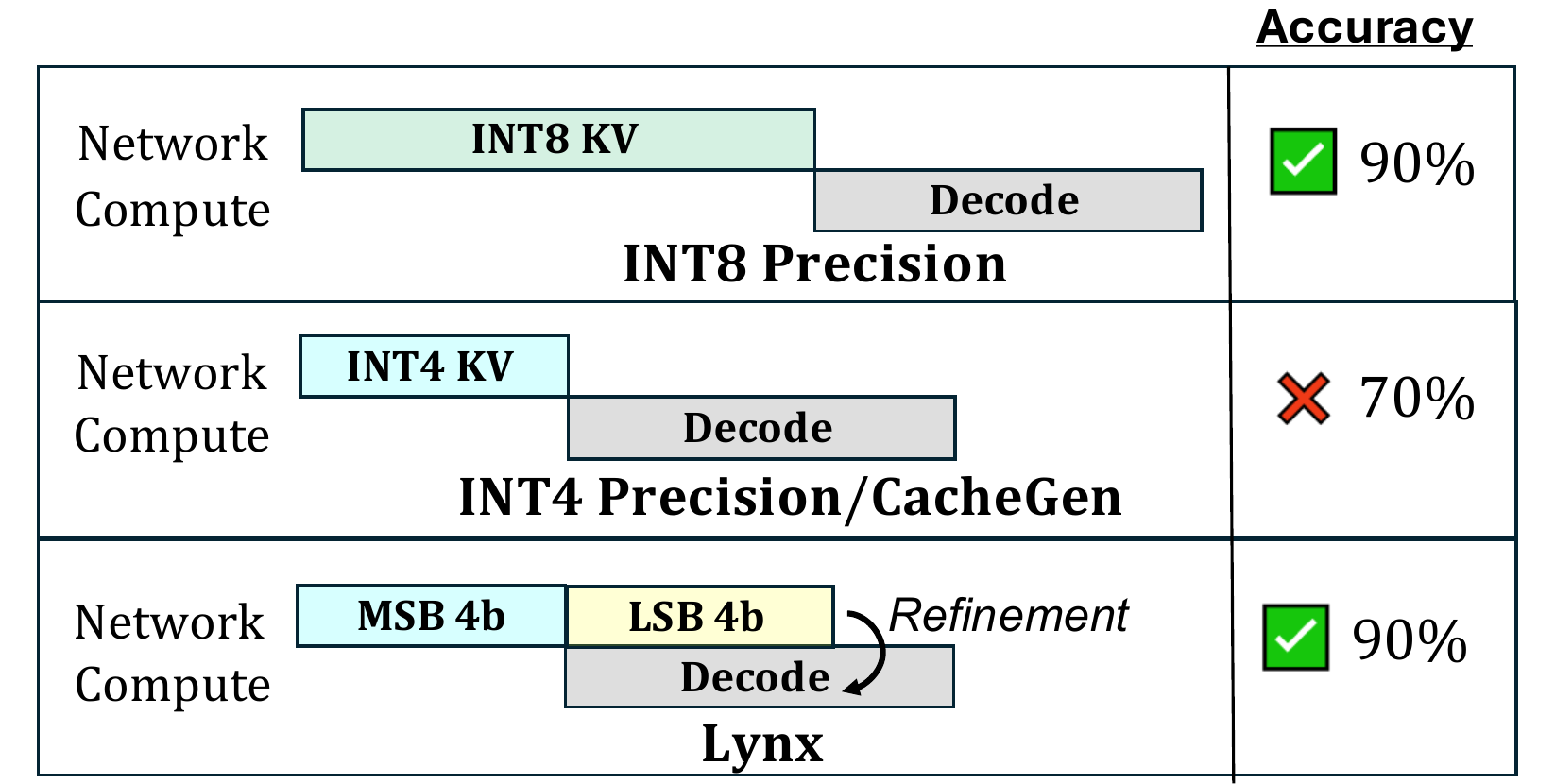}
    \caption{Inference timelines showing that \sysname achieves both high accuracy and low inference latency by overlapping the network communication of KV transfer with the computation of decoding.}
    \label{fig:sys-advantage}
\end{figure}

Existing systems mitigate this cost by quantizing the KV cache prior to transfer, which compresses KV cache~\footnote{Usually from BF16~\cite{kalamkar19bf16} or FP16.} to INT4 or INT8~\cite{hooper2024kvquant,liu2024kivi,liu2024cachegen,zhang2025hack}. This effectively reduces the KV volume, which yields accelerated KV transfer over the network. As illustrated in Figure~\ref{fig:sys-advantage}, while these approaches reduce data volume, they preserve a strict serialization barrier: the decode stage cannot begin until the full (compressed) KV cache has arrived and been reconstructed. Layer-wise pipelines partially overlap communication and computation, but when network bandwidth is the bottleneck, decoding still stalls at the boundary of each layer. As a result, current KV compression schemes reduce \textit{how much} data is transferred by not \textit{when} decoding can begin.

This paper challenges a core assumption underlying KV transfer: that the KV cache is an indivisible unit that must be fully received before it can be used. \textbf{Our key observation is that different bits within KV values contribute unequally to the attention probability distributions.} The Most Significant Bits (MSBs) determine the magnitude of attention scores and largely preserving the relative ranking of token associations, while the Least Significant Bits (LSBs) primarily refine precision. This asymmetry suggests that a partial KV cache is sufficient to begin the decoding phase, provided that the missing precision is recovered before final acceptance. 

Based on this insight, we develop \textbf{\sysname}, a system that implements progressive, bit-wise split-stream KV cache transfer and overlaps network communication with speculative decoding. We generalize the standard Draft-then-Verify paradigm~\cite{leviathan2023fast,chen2023accelerating}---which traditionally uses a smaller model to generate tentative tokens---to the network layer. In \sysname, the approximation is not derived from a separate model, but from the partial MSBs of the KV cache itself.

As shown in Figure~\ref{fig:sys-advantage}, \sysname splits the KV cache into two independent streams: a high-priority \textbf{Anchor Stream} containing the MSBs of quantized KV values, and a \textbf{Residual Stream} carrying the remaining precision. The Anchor Stream is transmitted first, allowing the decode instance to immediately begin attention computation and generate speculative tokens. While decoding proceeds, the Residual Stream is transfered in parallel and later used to verify and correct speculative outputs, ensuring that final generation recovers high-precision fidelity. 

To make this execution model practical, \sysname introduces a hierarchical quantization scheme that preserves attention probability distributions under aggressive MSB truncation, and a split-stream serving runtime that supports non-blocking KV access, and efficient verification. By orchestrating these components, \sysname transforms the network from a passive bottleneck into an active pipeline for speculative execution.
%Unlike prior speculative decoding approaches, Lynx does not rely on auxiliary draft model; instead it uses the same target model operating on a progressively refined KV cache. 

\begin{table}[t!]
\centering
\resizebox{\columnwidth}{!}{%
\begin{tabular}{@{}lcccccc@{}}
\toprule
\textbf{System} & \textbf{Quantization} & \textbf{Transfer Protocol} & \textbf{TTFT} & \textbf{TT32T} & \textbf{Quality} \\ \midrule
Standard (BF16) & None & Monolithic & $4.4$s & $5.9$s & $85.25\%$ \\ \midrule
INT8 & INT8 & Monolithic & $2.3$s & $3.8$s & $85.06\%$ \\ \midrule
INT4 & INT4 & Monolithic & $1.4$s & $2.9$s & $76.46\%$ \\ \midrule 
CacheGen~\cite{liu2024cachegen} & Delta Encoding & Monolithic & N/A & N/A & $80.07\%$ \\ \midrule
\rowcolor{gray!10} \textbf{Lynx (Ours)} & \textbf{Split-Stream} & \textbf{Pipelined} & $1.6$s & $3.4$s & \textbf{85.20\%} \\ \bottomrule
\end{tabular}
}
\vspace{0.2cm}
\caption{Comparison of Lynx against KV transfer baselines on the MMLU-Pro~\cite{wang2024mmlu} dataset, $16$K context length and Qwen 32B~\cite{qwen3} model. \sysname significantly reduces the Time-to-first-Token (TTFT) and Time-to-32-Token (TT32T) while maintaining BF16 generation quality. Full results are shown in \S\ref{sec:eval}.}
\label{tab:preview}
\end{table}

\begin{figure}[t!]
    \centering
    \centering
    \begin{minipage}[t]{0.9\linewidth}{
		\vspace{-0.00in}
		\begin{center}
		\includegraphics[width=\textwidth, ]{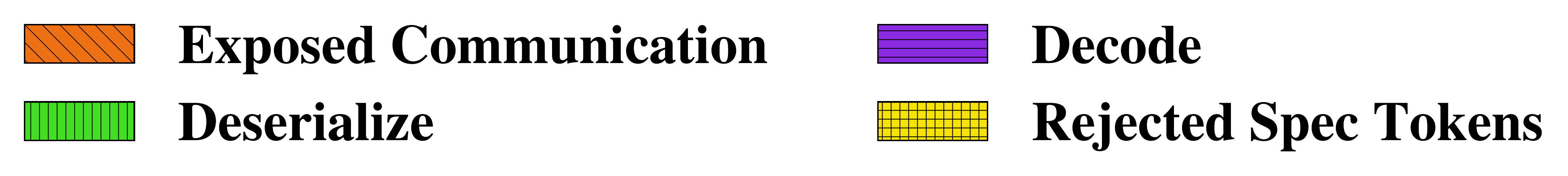}
		\end{center}
		}
        \end{minipage}

    \vspace{-0.3cm}
    \hspace{-0.2cm}
        \subfigure[LLaMA over MMLU-Pro, $128$K, $25$Gbps]{
		\begin{minipage}[t]{0.55\linewidth}{
		\vspace{-0.00in}
		\begin{center}
		\includegraphics[width=\textwidth, ]{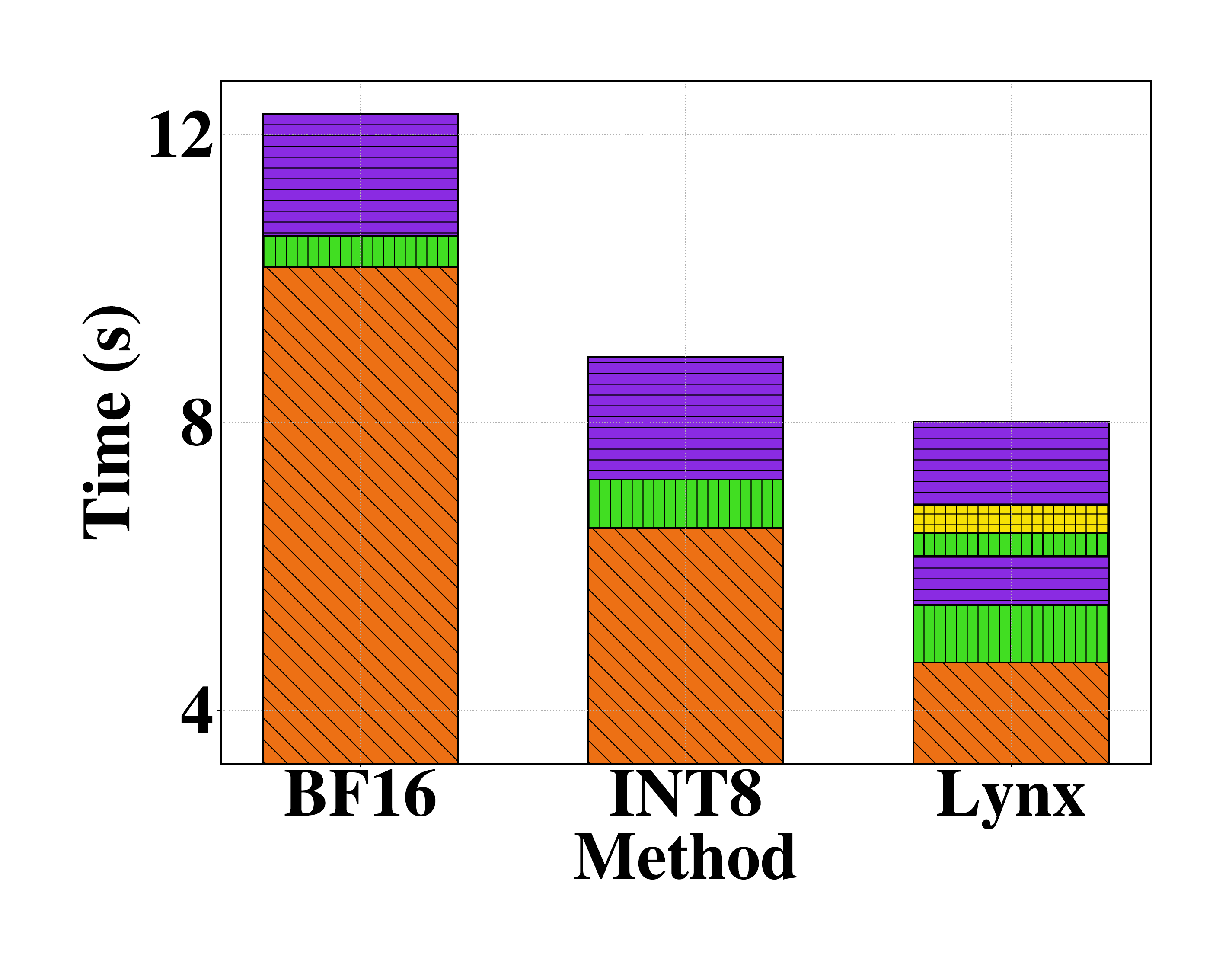}
		\end{center}
        \vspace{-0.1cm}
		}
		\label{subfig:bar-ttkt}
		\end{minipage}
	}

    \vspace{-0.2cm}
    %\caption{Breakdown of time consumed throughout our LLM inference pipeline with disaggregated prefill, as per the setups in Section~\ref{subsec:exp-scalability}. Bars for the prefill stage are omitted. The deserialize stage includes the overhead for decompression, and the time it takes to load serialized KV cache data into paged NPU memory.}
    \caption{Time breakdown of the disaggregated inference pipeline (Section~\ref{subsec:exp-scalability}). Prefill is omitted. 'Deserialize' includes decompression overhead and the loading of serialized KV data into paged device memory.}
    \label{fig:demo-decomposed-time}
\end{figure}

We evaluate \sysname on three different inference datasets with three different models including LLaMA~\cite{grattafiori2024llama}, Qwen~\cite{qwen3}, and Mistral~\cite{mistral24b}, and compare \sysname against state-of-the-art baselines including Cachegen's delta encoding~\cite{liu2024cachegen}\footnote{As our experiments are conducted on Ascend NPUs that CacheGen does not support yet, we only report the generation quality of our best-effort port of CacheGen.}, standard INT4 and INT8 quantization schemes. Our results show that \sysname achieves INT4-level KV transfer latency while maintaining INT8/BF16-equivalent inference accuracy, reducing both TTFT and Time-to-32-nd Token by up to $30.5\%$ and $10.7\%$ compared to the INT8 baseline and outperforms CacheGen's inference accuracy by up to $5.1\%$. Table~\ref{tab:preview} and Figure~\ref{fig:demo-decomposed-time} preview our results. These gains increase with longer context lengths and lower available bandwidth, precisely the regimes where disaggregated inference is most challenging. 
This paper makes the following core contributions: 
\begin{itemize}[leftmargin=*]
    \item  \textbf{Hierarchical Split-Stream Quantization:} We propose a quantization algorithm that physically partitions the KV cache into coarse-grained Anchors and fine-grained Residuals. By employing non-linear logarithmic quantization and outlier-aware chunking, we minimize the reconstruction error of the KV cache to preserve the attention output distribution within a constrained bit budget.
    %\item \textbf{Hierarchical Split-Stream Quantization:} We propose a quantization algorithm that physically partitions the KV cache into coarse-grained Anchors and fine-grained Residuals. By employing non-linear logarithmic quantization and outlier-aware chunking, we minimize the reconstruction error of the Anchor stream to preserve the approximate attention output distribution within a constrained bit budget.
    \item \textbf{Split-Stream Pipelining Architecture:} We propose a network-system co-design 
    %We design a network-system co-design 
    that prioritizes the transmission of critical Anchor bits. This architecture integrates a custom SerDes protocol with a non-blocking serving runtime, enabling the Decode phase to commence on partial KV states while the Residual stream is transferred concurrently in the background.
    \item \textbf{Speculative Decoding Verification:} We design a verification protocol leveraging the formal guarantees of speculative decoding. This ensures that the distribution of tokens generated by our inference system is identical to those drawn from the un-quantized full-precision model, guaranteeing the same expected generation quality.
    %that refines the output distribution as residual bits arrive. T
    %his guarantees that the speculative draft converges to the exact mathematical equivalent of the full-precision baseline, ensuring zero degradation in generation quality.
\end{itemize}

\section{Background \& Limitations}
Large Language Model (LLM) services are built upon the Transformer architecture~\cite{vaswani2017attention}, which relies fundamentally on the Attention mechanism operating at each layer. The standard inference process is \textit{autoregressive}: the model takes an input sequence (the prompt), and generates output tokens sequentially, where each new token is conditioned on the entire previous context. 

During each generation step, the attention mechanism at each layer must compute the relationship between the current input token, and all preceding tokens. Formally, for the current token, the model generates three projections: the Query ($\mathbf{Q}$), Key ($\mathbf{K}$), and Value ($\mathbf{V}$) tensors. The current $\mathbf{Q}$ is then used to calculate an attention output $O$ against the $\mathbf{K}$ and $\mathbf{V}$ tensors of all previous tokens defined as:

\begin{equation*}
    O=\text{softmax}(QK^T/\sqrt{d_k})\cdot V,
\end{equation*}
where $d_k$ is the model head dimension.  %the attention output for the last layer is used to generate the output tokens $o_t$. 
To avoid redundant re-computation, these historical Key and Value tensors are preserved in memory as the \textit{KV Cache}. This structure typically resides in high-bandwidth memory (HBM) on the device accelerators, or is offloaded to host DDR to support long-context generation~\cite{rasley2020deepspeed,chen2025impress,liu2025lmcache}.

\subsection{Prefill, Decode, and Disaggregation}
% LLM Inference naturally divides into two distinct phases with differing computational characteristics~\cite{agrawal2023sarathi,jiang2024minference}. 
% \begin{enumerate}
%     \item \textit{Prefill (Compute-Bound):} The model processes the user's entire input prompt to populate the initial KVCaches and generate the first output token. This phase is highly parallelizable and demands high computational requirements.
%     \item \textit{Decode (Memory-Bound):} The model generates output tokens sequentially based on previous tokens. This phase is limited by memory bandwidth, as the growing KV cache must be loaded from HBM to on-chip SRAM for every single token generated~\cite{pope2023efficiently,kwon2023efficient}. 
% \end{enumerate}

% To optimize resource utilization and mitigate resource contention between these phases, recent systems adopt \textit{Disaggregated Inference} (PD-Disaggregation)~\cite{patel2024splitwise,zhong2024distserve,jin2024p}. In this paradigm, prefill requests are assigned to a distinct set of instances, while decode requests are routed to another. While this separation improves throughput and reduces interference, it transforms the KV cache into a critical network payload that must be transferred between instances.
Standard LLM inference partitions execution into two phases with distinct hardware requirements: a compute-intensive \textit{Prefill} phase that processes the prompt in parallel, and a memory-bandwidth-bound \textit{Decode} phase that generates tokens sequentially~\cite{agrawal2023sarathi,jiang2024minference}. To address the resource contention between these phases, state-of-the-art systems adopt \textit{Disaggregated Inference} (PD-Disaggregation), routing prefill and decode requests to specialized instances~\cite{patel2024splitwise,zhong2024distserve}. 
\\While this architecture optimizes throughput, it creates a rigid dependency: the decoding instance cannot begin computation until it receives the requisite KV cache data from the prefill instance, whether transferred as a monolithic or in layer-wise chunks.

%\william{I think section 2 up to this point is good. However, you introduce prefill and decode only to then discuss continous batching and then reintroduce p/d disagg. Perhaps this could be restructured. It is not clear whether continuous batching even needs to be introduced in this formal way since this isn't the setting we care about. }

\subsection{The KVCache Transfer Bottleneck}
\label{sec:kvbottleneck}
Transferring the KV cache over the network creates a scalability barrier. The memory footprint grows linearly with sequence length ($L$). For a model with hidden dimension $D$ (product of attention heads and head dimension) and layer count $N_{layers}$, the total KV cache size is defined as:
\begin{equation}
    \text{Size}_{KV} = K_{factor} \cdot L \cdot D \cdot N_{layers} \cdot \textit{Precision} 
\end{equation}
where $K_{factor}$ is 2 for standard attention and 1 for Multi-Head Latent Attention. %In modern long-context scenarios (e.g. 256k to 1M tokens), this state can reach tens of gigabytes per request depending on the model architecture~\cite{vonplaten2024llama31,hooper2024kvquant}. 
Consider a single request with a 128k context window processed by Qwen3-235B-A22B ($N_{layers}=94, H_{kv}=4, D_{head}=128$). This single request generates approximately 23.5 GB of KV data in bfloat16\footnote{Calculation: $2 \times 94 \times 131,072 \times 4 \times 128 \times 2 \approx 23.5 \text{ GB}.$}. 
Transferring this payload creates a massive latency penalty in two primary deployment scenarios. 
\\ \textbf{Scenario 1: Disaggregated Prefill-Decode.} On a standard 100Gbps TCP/IP interconnect, transferring this 23.5 GB payload requires roughly 2 seconds, effectively stalling the inference pipeline. While state-of-the-art engines~\cite{kwon2023efficient,nvidia2025dynamo} attempt \textit{layer-wise pipelining} (computing layer $i$ while receiving $i+1$), this approach fails when the transfer over the network exceeds computation time ($T_{comm} > T_{comp}$), negating the benefits of disaggregation.
\\ \textbf{Scenario 2: Remote Context Retrieval.} Similarly, serving long-context models often requires retrieving KV cache states from remote storage~\cite{liu2025lmcache}. Even in high-performance clusters equipped with 400Gbps, the transfer consumes over 0.5 second \textit{per request}. Standard inference engines operating in blocking modes suffers a direct penalty on the TTFT proportional to the network latency.
\\Consequently, simply increasing physical bandwidth is insufficient. Reducing the data volume itself is required, leading to two primary strategies: \textit{quantization} and \textit{compression}.
%On a standard 100Gbps TCP/IP interconnect, this transfer requires roughly 4 seconds, effectively stalling the inference pipeline and negating the latency benefits of PD Disaggregation~\cite{patel2024splitwise}. Even in high-performance clusters equipped with 400Gbps InfiniBand/RDMA, the transfer can consumes over 1 second per request. When scaling to multi-million token contexts regime~\cite{lin2024infinite,meta2025scaling}, the frequency of these transfers saturates the physical link capacity. Consequently, simply increasing physical bandwidth is insufficient. Reducing the data volume itself can alleviate the bandwidth challenge, there are two potential complementary strategies: \textit{quantization} and \textit{compression}.

\subsection{Quantization}
\label{sec:quantizationandcompression}
Quantization is an effective method to reduce data volume. In LLMs, applying quantization to both weights and activations has proven highly effective for reducing memory consumption and accelerating inference~\cite{han2016deepcom,jacob2018quant,hooper2024kvquant,xiao2023smoothquant,zhao2024atom,lin2025qserve}. Unlike weight quantization that is often handled pre-deployment, KV cache quantization is primarily handled at runtime, where each request corresponds to unique data.

%\subp{Standard \& Grouped Quantization.} The objective is to map high-precision floating-point values to lower-precision integers. For a tensor $\mathbf{X}$, the quantization process is formally defined as:
%\begin{equation}\label{eq:quant}
%    Q(\mathbf{X}) = \left\lfloor \frac{\mathbf{X}}{S} + Z \right\rceil \in [q_{min}, q_{max}]
%\end{equation}
%The scaling factor $S$ and zero-point $Z$ are derived from the min-max range $[\mathbf{x}_{min}, \mathbf{x}_{max}]$. $[q_{min}, q_{max}]$ is the target integer range $X$ is quantized into. For example, with standard INT4 quantization $q_{min}=-8$ and $q_{max}=7$.
%To improve precision, \textit{Grouped Quantization} partitions the tensor into smaller blocks, calculating independent scale factors for each. This isolates the range of local values, preventing global extremes from skewing the scale for the entire tensor.

\subp{Standard \& Grouped Quantization.}
Standard quantization maps floating-point values to integers via a linear transform: $Q(\mathbf{X}) = \lfloor \mathbf{X}/S + Z \rceil$. Critical to this process is the scaling factor $S$, which is derived from the tensor's dynamic range $[\mathbf{x}_{min}, \mathbf{x}_{max}]$.
However, relying on a global scale $S$ for an entire tensor is proven to be insufficient~\cite{dettmers2022llm,xiao2023smoothquant}. Grouped Quantization~\cite{dettmers2022llm,zhao2024atom} further partitions the tensor into smaller blocks, calculating independent scales for each block to capture finer grained range of values. 

\subp{The Challenge of Activation Outliers.} Despite these techniques, existing quantization schemes faces a critical hurdle in LLMs: activations are dominated by extreme outliers~\cite{dettmers2022llm,wei2022outlier}. For a structure such as KV cache, the hidden dimension $D_{head}$ consists of independent feature columns, referred to as \textit{channels}. Empirical analysis shows that a small subset of features consistently holds values up to $100\times$ larger than the rest. 
These outliers break linear mapping with the standard quantization schemes because the scale $S$ must be expanded to accommodate the extreme magnitude $m$. For a non-outlier channel $i$ with local maximum $m_i$ and global maximum $m$, the \textit{effective number of bins} is reduced to:
\begin{equation}
    \text{Bins}_{eff} \approx 2^B \cdot \frac{m_i}{m}
\end{equation}
where $2^B$ is the total bucket count. With a $100\times$ outlier ratio ($m \approx 100 m_i$), the resolution for the majority of non-outlier channels collapses. In an 8-bit system (256 bins), this effectively degrades precision to roughly 1.5 bits ($\log_2 3$), causing large quantization error~\cite{xiao2023smoothquant}.

While quantization effectively mitigates the transfer bottleneck by reducing the bit-width of the KV cache, it is inherently lossy. Compressing the attention keys and values introduces approximation errors that accumulate during the autoregressive generation process. 
%Consequently, this leads to a degradation in generation quality compared to the full-precision baseline.  
Relying solely on quantization therefore forces a compromise: one must sacrifice output quality to achieve lower startup latencies. A standard technique for resolving this trade-off—by allowing approximate proposals that are later verified without altering the final output distribution—is speculative decoding~\cite{leviathan2023fast,chen2023accelerating}. %gingfung love this
%A standard technique for resolving this trade-off is speculative decoding, which enables low-latency approximate outputs to be proposed and subsequently verified without altering the final output distribution~\cite{leviathan2023fast,chen2023accelerating}.

\subsection{Speculative Decoding} \label{sec:specdec}
Speculative decoding accelerates LLM inference~\cite{leviathan2023fast,chen2023accelerating}. It employs a lightweight approximation of the target model (often referred to as a draft model) to rapidly generate a sequence of candidate tokens. The target model then verifies the candidates in parallel, accepting a prefix of the sequence and correcting where the approximation diverges. 

A defining property of speculative decoding is its losslessness: it guarantees that the final output distribution is mathematically equivalent to sampling from the target model directly. The acceptance rate of speculative decoding depends on the alignment between the draft and target distributions; if the draft model is a poor approximation, frequent rejections occur, thereby diminishing the efficacy of the approach. Consequently it is critical that the draft model effectively approximate the target model.

% Given a draft model $q$ and a target model $p$ a draft token drawn from the draft model $\Tilde{x}\sim q(\cdot)$ is accepted with probability 

% \begin{align}\label{eqn:spec_dec}
%     P(\text{accept}) = \min\left( \frac{p(\Tilde{x})}{q(\Tilde{x})}, 1\right)
% \end{align}
% \begin{align}\label{eqn:spec_dec}
%     P(\text{accept}) = \mathbf{1}\left( \tilde{x} = \operatorname*{argmax}_x p(x) \right)
% \end{align}

\section{Not All Bits Are Equal}
This section proposes a new perspective on KV availability: initiating decoding prior to complete KV cache transmission. We argue that while early decoding is viable, standard linear quantization is inadequate to support it.
% This Section motivates the 
% rethinks KV availability. We begin by asking whether decoding can commence before arrival of the full KV cache, going on to argue why standard linear quantization is insufficient to support it.

\subsection{Rethinking KV Availability} Current systems suffer from a "transfer bottleneck" because they view KV cache availability as a binary state. We argue for a shift toward \textit{progressive execution}, analogous to progressive rendering in web browsing. In this model, generation commences using a low-precision subset of the KV cache. These approximated outputs are then validated upon the arrival of the remaining precision bits. 

Hypothetically, initiating decoding with only 50\% of the data volume could halve the TTFT. However, this introduces two challenges. First, a formal verification protocol is required to guarantee the accelerated output remains identical to the standard, full-precision model. Second, to ensure these approximate outputs actually \textit{pass} validation, it is critical to prioritize the transmission of the most information-dense segments of the cache. While a naive method might simply transmit Most-Significant Bits (MSBs), as we show, a more robust quantization strategy is required to minimize verification failures and fully realize these latency gains.

% \begin{figure}
%     \centering
%     \includegraphics[width=0.7\linewidth]{exp_figures/theory_ntokens.pdf}
%     \caption{Extra number of tokens a model can theoretically decode by overlapping the network transmission of the 4 LSBs with computation~\footnote{The number of tokens is computed as the ratio of the transmission time of the LSBs of the KV cache and the token generation speed in TPOT of the model.}} 
%     \label{fig:theory_benefits}
% \end{figure}

\begin{figure}[t!]
    % \vspace{0.38in}
    % \hspace{-0.3cm}
        
        \subfigure[Attention output]{
		\begin{minipage}[t]{0.375\linewidth}{
		\vspace{-0.3cm}
		\begin{center}
		\includegraphics[width=\textwidth, trim=2cm 2cm 2cm 2.0cm]{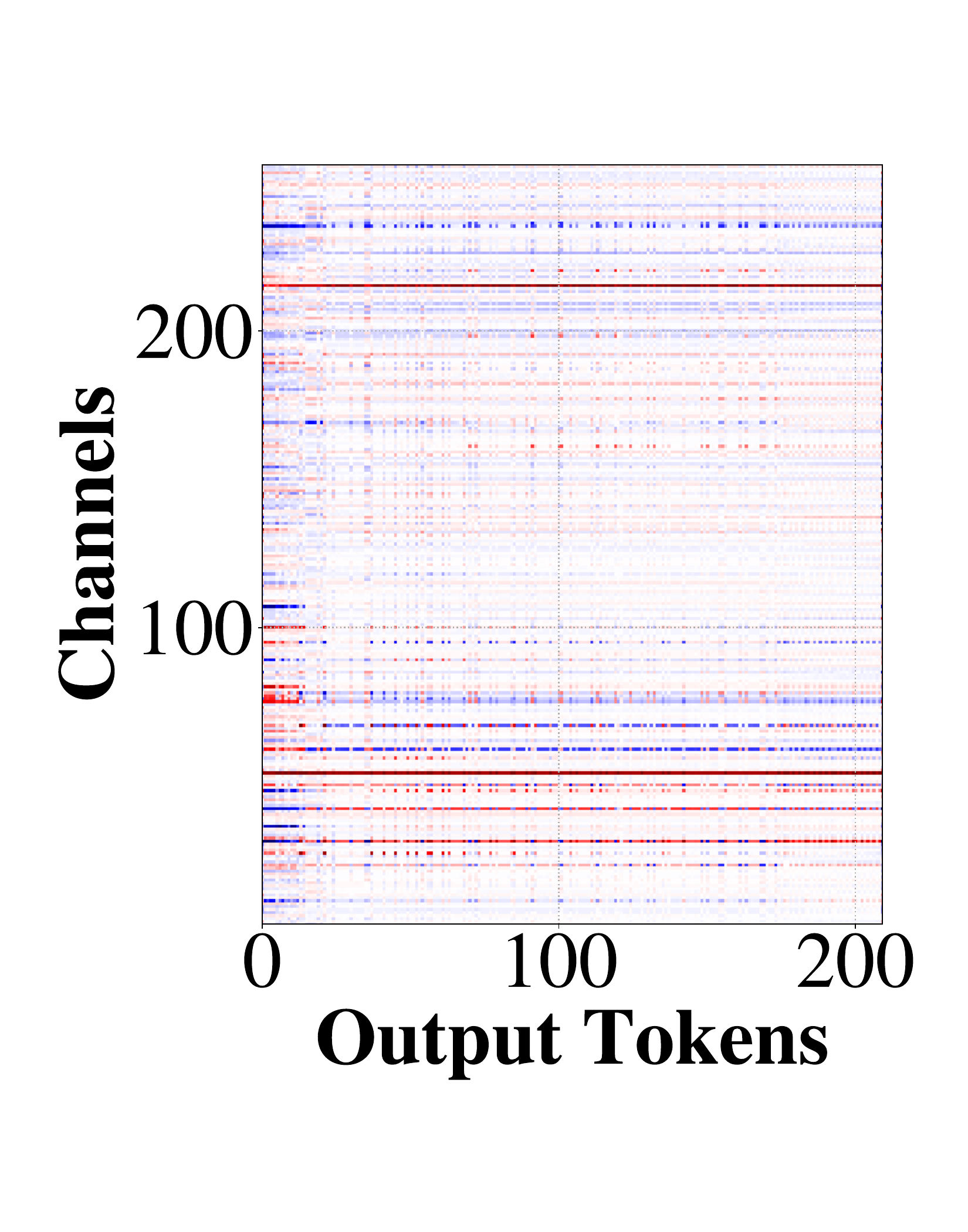}
                \vspace{-0.5cm}
		\end{center}

		}
		\label{subfig:attn_out}
		\end{minipage}
	}
        \subfigure[Compression error of INT4]{
		\begin{minipage}[t]{0.48\linewidth}{
		\vspace{-0.2cm}
		\begin{center}
		\includegraphics[width=\textwidth, trim=2cm 2cm 2cm 2.0cm]{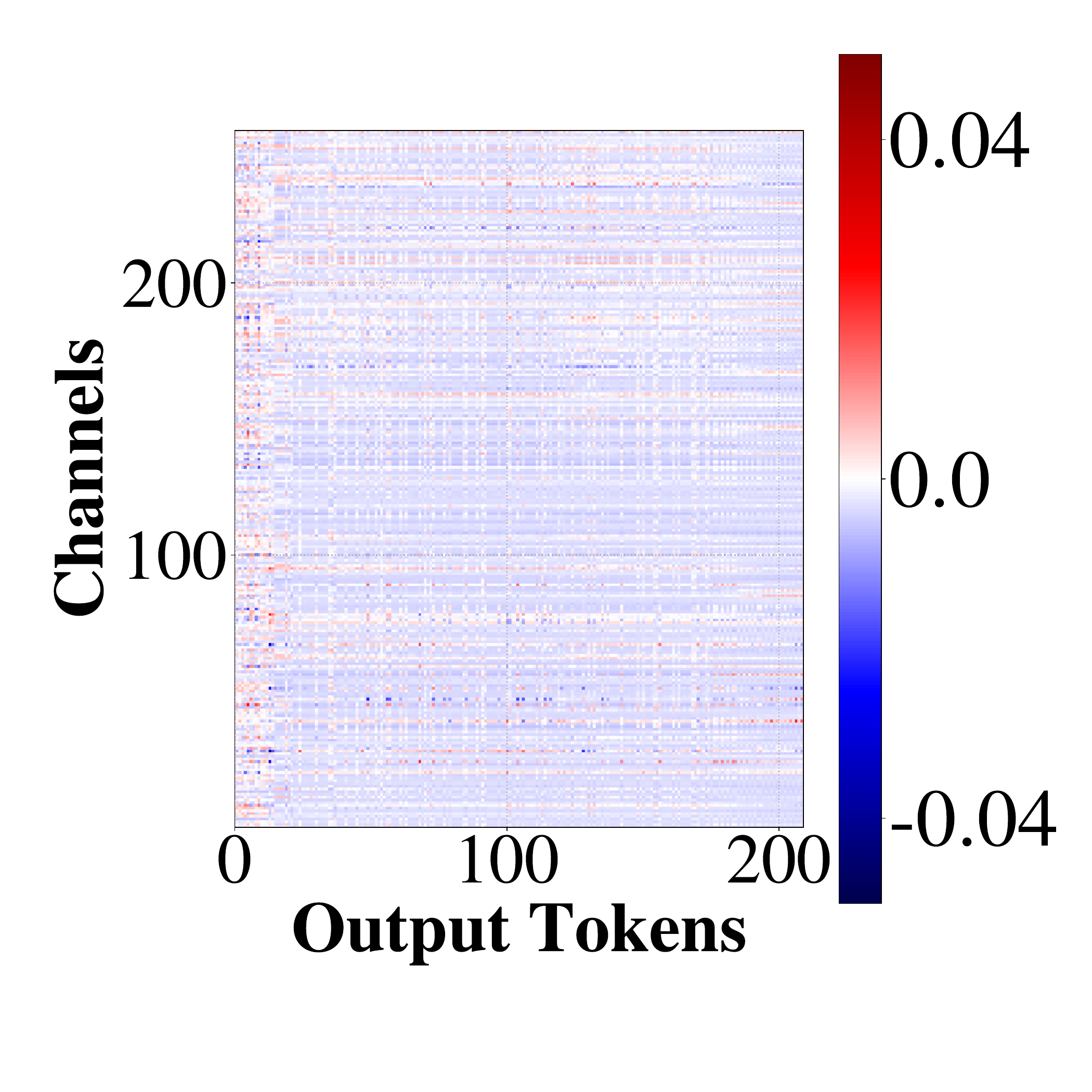}
                \vspace{-0.5cm}
		\end{center}

		}
		\label{subfig:kv_int4_residual}
		\end{minipage}
	} 
    %
    % \hspace{-0.3cm}

    % \gingfung{We think that we should only mention our MSB approach in evaluation}
  %   \hspace{-0.1cm}
  %       \subfigure[Compression error of \sysname with MSB]{
		% \begin{minipage}[t]{0.36\linewidth}{
		% \vspace{-1.25in}
		% \begin{center}
		% \includegraphics[width=\textwidth, ]{exp_figures/attention_outputs_comparison_est.pdf}
		% \end{center}
  %       \vspace{-0.1cm}
		% }
		% \label{subfig:kv_msb_residual}
		% \end{minipage}
  %   }
  \vspace{-0.3cm}
  \caption{Visualization of the attention output (left) and the corresponding INT4 quantization error (right). The error magnitude is significant relative to the feature scale, indicating substantial precision loss.}
    \vspace{-0.1cm}
    % \caption{The attention output without KV quantization (left), compared with the compression error of the attention output with INT4 (right). The compression error is fairly large.} %In (c) the attention computation is based on the most significant 4-bit (MSB) of \sysname's quantized KV cache of the context tokens.}
    \label{fig:kv_error}
\end{figure}

\begin{figure}[t!]
    \centering
    % \vspace{0.38in}
    \hspace{-0.3cm}
        \subfigure[$K$ Cache, Layer 0]{
		\begin{minipage}[t]{0.45\linewidth}{
		\vspace{-1.4in}
		\begin{center}
		\includegraphics[width=\textwidth, ]{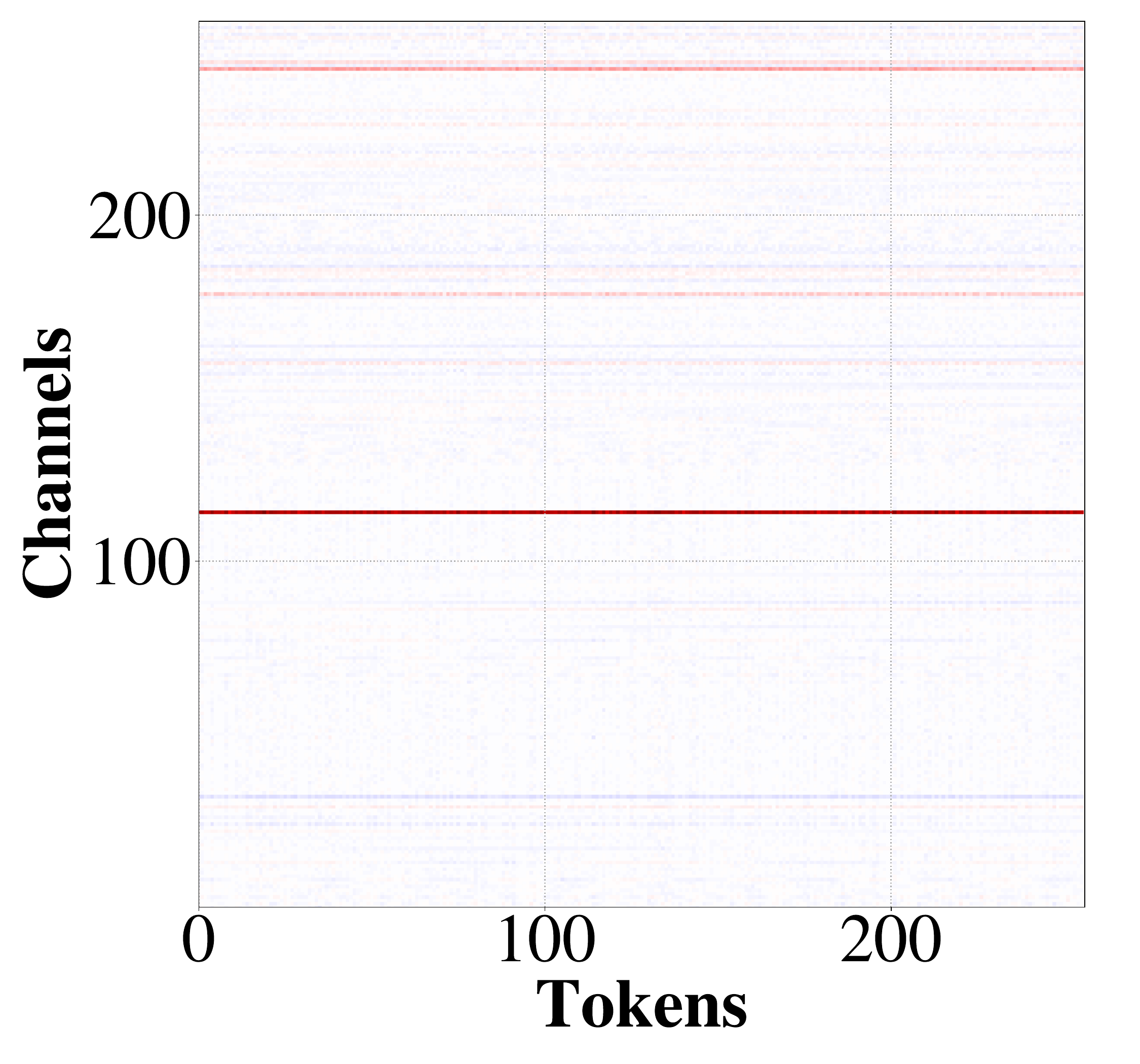}
		\end{center}
            \vspace{-0.1cm}
		}
		\label{subfig:heatmap_0}
		\end{minipage}
	}
    \hspace{-0.3cm}
        \subfigure[$K$ Cache, Layer 32]{
		\begin{minipage}[t]{0.55\linewidth}{
		% \vspace{-0.38in}
		\begin{center}
		\includegraphics[width=\textwidth, ]{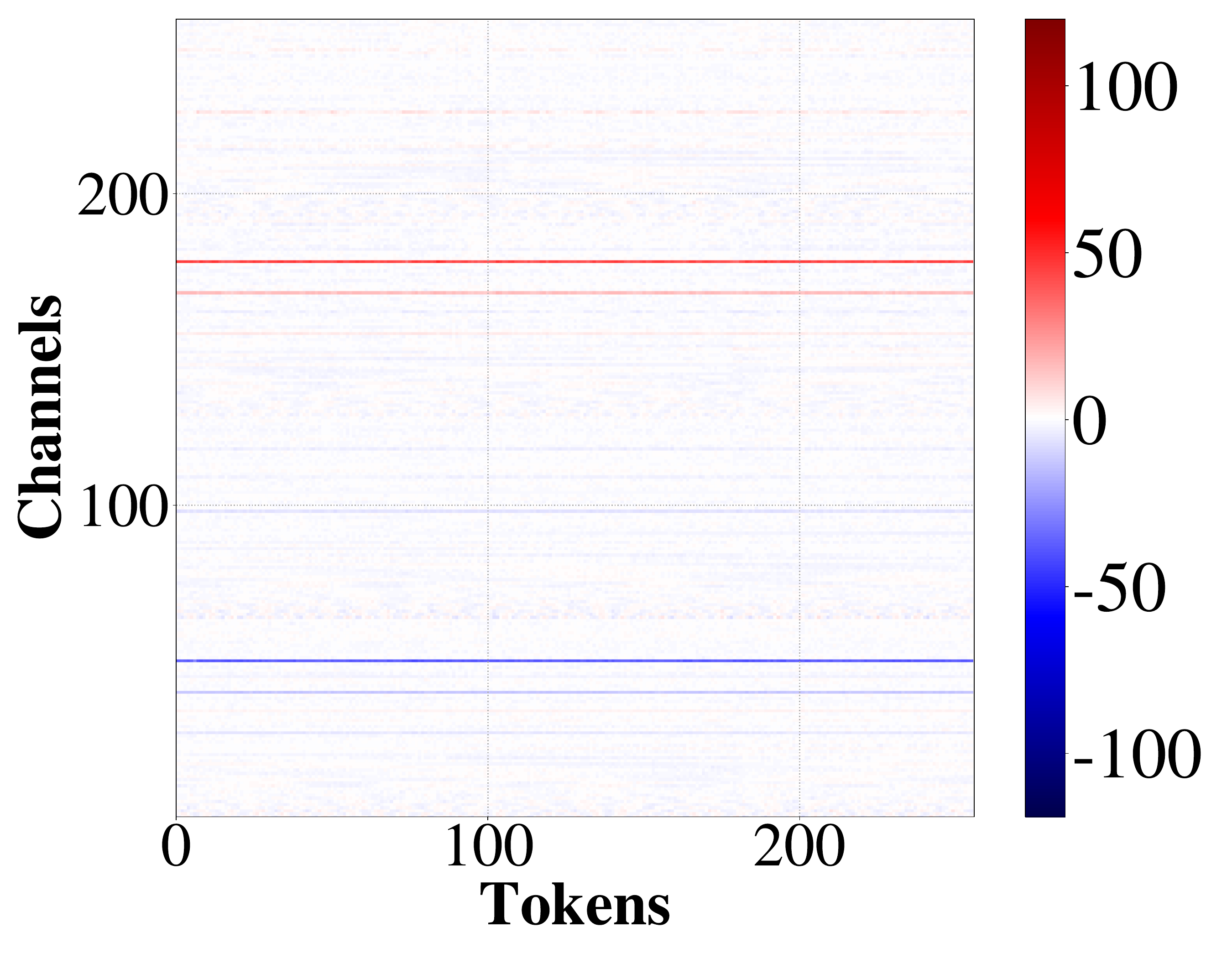}
		\end{center}
        \vspace{-0.1cm}
		}
		\label{subfig:heatmap-32}
		\end{minipage}
	}
    \vspace{-0.3cm}
    \caption{Visualization of KV cache values at different layers (Qwen 32B on the MMLU-Pro dataset). %The x-axis and y-axis correspond to token and channel dimensions, respectively. 
    The persistence of extreme values in specific outlier channels (visible as dark horizontal stripes) highlights the need for a per-channel quantization approach.}
    % \caption{KV cache data distribution over different layers running Qwen 32B model over the MMLU-Pro dataset. The x-axis and the y-axis are tokenwise and channelwise dimensions respectively. We note that the extreme outliers concentrate within a few channels throughout the tokens, which motivates our per-channel quantization design.}
    \label{fig:kv-cache-distribution}
    \vspace{-0.0cm}
\end{figure}

\begin{figure}
    \centering
    \includegraphics[width=0.6\linewidth]{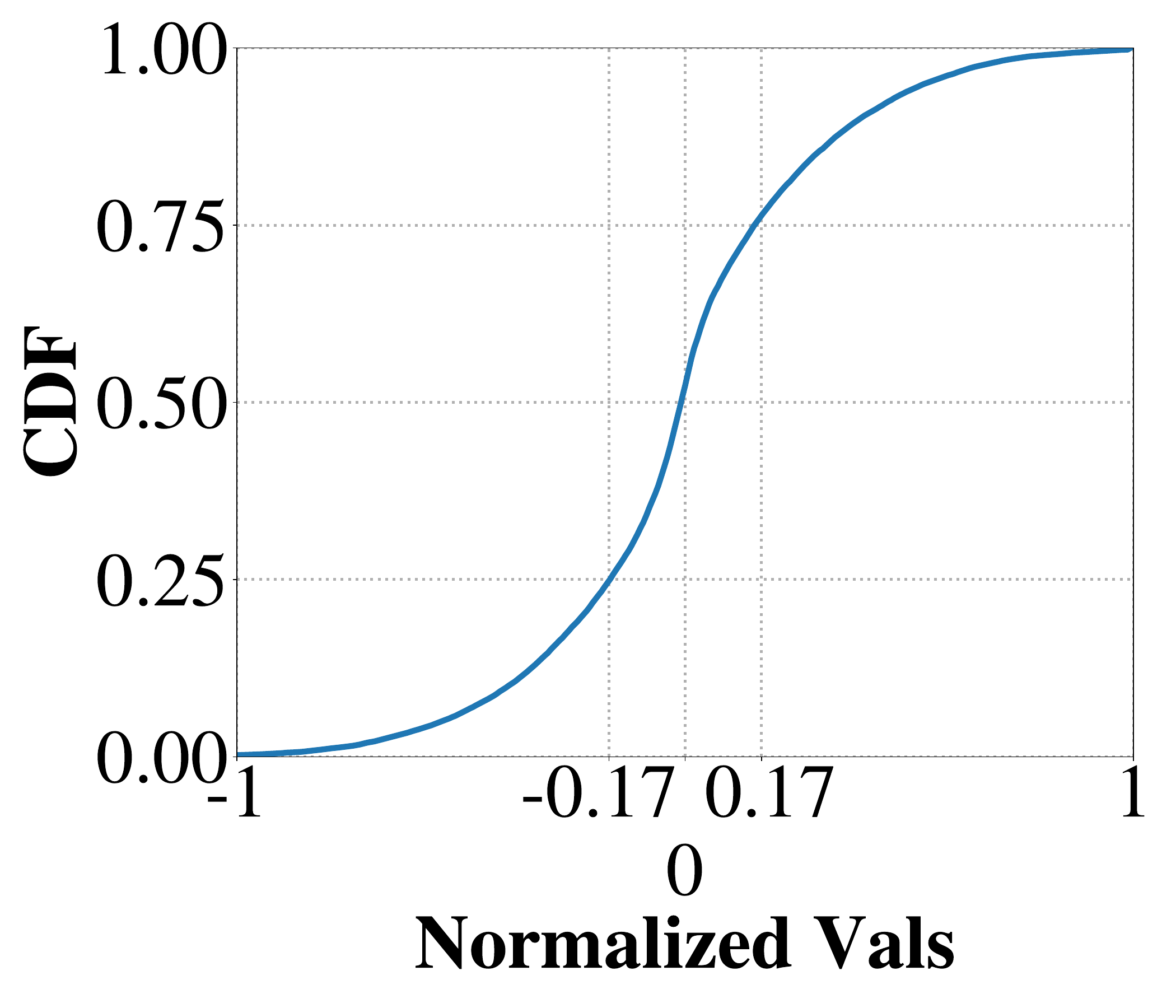}
      \vspace{-0.2cm}
    \caption{The KV cache data distribution after tokenwise (per-channel) normalization that effectively isolates extreme outliers. The x-axis is normalized values from $-1$ to $1$ and the y-axis is the cumulative proportion of the KV cache data whose normalized value is smaller than the given value. We see that $50\%$ of the KV cache elements concentrate within the middle $17\%$ of the value range.}
    \label{fig:kdist_around_0}
\end{figure}

\subsection{The Limitation of Linear Quantization}
\label{subsec:thelimitoflinear}
%\gingfung{here I am trying to say 1) standard int4 does not work, 2) because of linear mapping, 3) motivate the need of a non-linear approach}

We first examine the feasibility of standard low-precision quantization. As shown in Figure~\ref{fig:kv_error}, standard INT4 quantization of the KV cache significantly distorts the attention output distribution compared to the full-precision ground truth.
This degradation is largely attributable to the outliers discussed in Section~\ref{sec:quantizationandcompression}. ~\Cref{subfig:heatmap_0} and~\Cref{subfig:heatmap-32} reveal that certain feature dimensions exhibit consistently high magnitudes. Accommodating these extremes forces the quantization grid to stretch, resulting in a loss of fidelity for non-outlier elements.
% This degradation is primarily driven by the outlier-induced quantization ranges discussed in Section~\ref{sec:quantizationandcompression}. This is because the quantization range must be stretched to accommodate the extreme values in the outlier channels, as shown in ~\Cref{subfig:heatmap_0} and ~\Cref{subfig:heatmap-32}.

However, simply removing outliers does not solve the problem. As illustrated in \Cref{fig:kdist_around_0}, even after isolating the extremes, the remaining KV cache values do not spread uniformly. Instead, they cluster densely around zero, following a sharp Laplacian-like distribution~\cite{NEURIPS2022_5e07476b}. Applying a linear mapping to this non-uniform data creates two  inefficiencies:
\begin{itemize}[leftmargin=*]
    \item \textbf{Wasted Capacity:} A significant portion of the quantization bins are allocated to the tail ends of the distribution where almost no data points exist.
    \item \textbf{Starved Precision:} The zero center, which contains the majority of the information, is forced to share a small number of bins, causing severe information loss.
\end{itemize}

These observations show that KV transfer need not be a blocking atomic operation. Decoding can begin using a partial, low-precision view of the KV cache, but doing so requires careful control over approximation error to ensure that speculative outputs remain faithful to the full-precision model. In the next Section we translate these insights into concrete system design choices for \sysname, which enables progressive KV transfer through prioritized streaming and lossless speculative verification.

\begin{figure}[t!]
    \centering
    \vspace{-0.1cm}
    \includegraphics[width=0.9\linewidth]{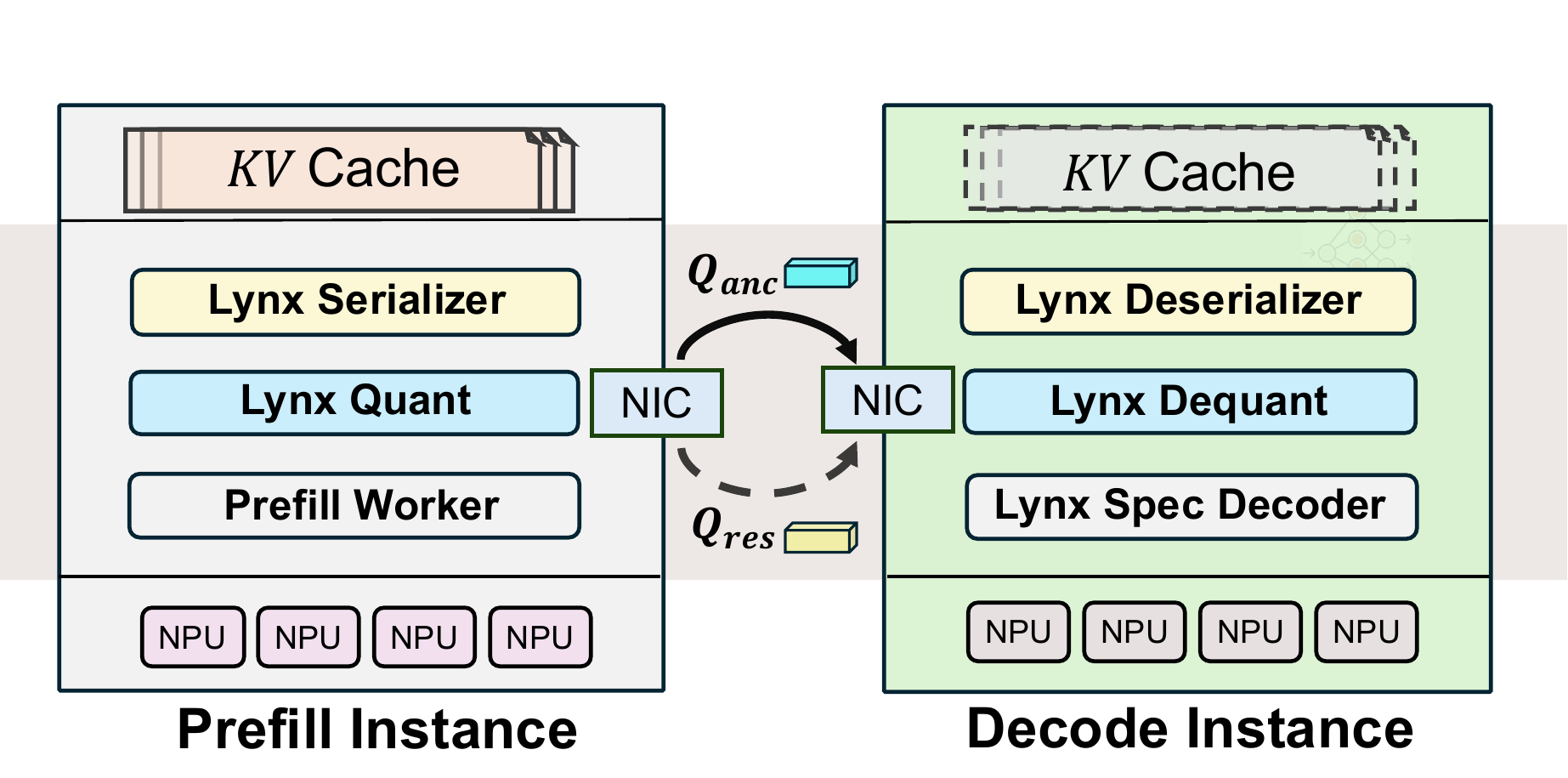}
    \vspace{-0.1cm}
    \caption{System overview of \sysname.}
    \label{fig:sys-overview}
\end{figure}

\section{Lynx: Enabling Progressive Execution}\label{sec:design}

\iffalse
\begin{table}[t]
    \centering
    \resizebox{1\linewidth}{!}{
    \begin{tabular}{|c||l|} \hline
        \textbf{Notation} & \textbf{Meaning} \\ \hline \hline
        MSB-4b & The  \\ \hline
        LSB-4b & \\ \hline
    \end{tabular}
    }
    
    \vspace{0.0cm}
    \caption{Notations that are commonly used in the paper.}
    \vspace{0.1cm}
    \label{tab:notations}
\end{table}
\fi

%\noindent\textbf{Implication for System Design.} This analysis necessitates a bifurcated approach. To sustain high-fidelity attention output with quantization schemes, a serving system must (1) explicitly preserve outlier channels to maintain the output distribution, and (2) apply a \textbf{non-linear transformation} to the data to maximize the information density of the remaining bits.

%To address the serialization bottleneck, 
% \gingfung two key components across Sender / Receiver. three main working mechanism ***
We present \textit{\sysname}, a system designed to enable progressive execution during KV transfer. \sysname, depicted in \Cref{fig:sys-overview}, is built upon a non-blocking runtime consisting of two main components: i) a Stream Serializer/Deserializer (SerDes) for high-throughput data handling, and ii) a Speculative Decoder for asynchronous generation. By coupling this streaming runtime with a distribution-aware quantization algorithm, \sysname effectively hides network transfer latency behind the useful work of speculative token generation.

\sysname redefines the contract between prefill and decode instances. Instead of treating the KV cache as a monolithic tensor that must be transferred atomically, \sysname manages it as a bifurcated data stream. The system operates via three pipelined mechanisms:
%As illustrated in Figure~\ref{fig:workflow-matrix}, the system operates in three pipelined stages:

\begin{enumerate}[leftmargin=*]
    \item \textbf{Hierarchical Quantization:} The prefill instance employs a novel hierarchical quantization algorithm to decompose the KV cache into two logical components: i) a high-fidelity Anchor Stream, containing the MSBs and outlier-aware scalars, and ii) a Residual Stream, containing the LSBs correction terms.
    \item \textbf{Prioritized Transmission:} \sysname prioritizes the network transfer of the Anchor Stream, minimizing the time-to-first-token for the decode instance.
    \item \textbf{Speculative Generation:} Upon receiving the Anchor Stream, the decode instance immediately commences generation, using the approximate KV cache to produce draft tokens. Once the Residual Stream arrives, the instance verifies the draft tokens, confirming or possibly correcting the output.
\end{enumerate}

% To orchestrate this workflow, as depicted in \Cref{fig:sys-overview}, \sysname is built upon a non-blocking runtime consisting of two main components: 1) a Stream Serializer/Deserializer (SerDes) for high-throughput data handling, and 2) a Speculative Decoder for asynchronous generation.

% By coupling this streaming runtime with a distribution-aware quantization algorithm, \sysname effectively hides network transfer latency behind the useful work of speculative token generation.
The following subsections detail the design of the three core mechanisms \sysname: Hierarchical Quantization algorithm (Section~\ref{sec:hierarchical_quant}), the Prioritized Transmission mechanism (Section~\ref{sec:bit_stream_pipe}), and the Verification mechanism (Section~\ref{sec:verification}).

\subsection{Hierarchical Quantization}
\label{sec:hierarchical_quant}
As discussed in Section~\ref{subsec:thelimitoflinear}, standard quantization relies on a linear mapping using a global or per-channel scaling factor $S$. 
However, applying directly to the KV cache is insufficient: the range of outlier channels forces $S$ to be large, collapsing the small values into the zero bin ($Z$).

To overcome these drawbacks, we propose a \textit{Hierarchical Non-Linear quantization scheme}. We introduce two key modifications to the standard linear quantization:
\begin{itemize}[leftmargin=*]
    \item \textbf{Granularity Shift:} Instead of a single scale $S$, we decompose the normalization into a two-level channel-wise hierarchy (Page $\gamma$ and Chunk $\sigma$) to preserve and isolate the outliers spatially.
    \item \textbf{Non-Linearity:} We replace the uniform linear mapping with a non-linear logarithmic transformation. This aligns the quantization bins with the distribution of the residuals, allocating higher precision to the dense region to maximize the information carried by each bit.
\end{itemize}

\begin{algorithm}[t]
\caption{Lynx Hierarchical Quantization (Vectorized)}
\label{alg:quantization}
\begin{algorithmic}[1]
\Require Page Block $\mathbf{X} \in \mathbb{R}^{H \times P}$ ($H$: Block Size, $P$: Tokens)
\Require Chunk Size $C$
\Ensure Anchor $\mathbf{Q}_{anc}$, Residual $\mathbf{Q}_{res}$

\vspace{0.1cm}
% \State \textit{// Stage 1: Page-Level Normalization}
% \State $\boldsymbol{\gamma}_{min} \leftarrow \min(\mathbf{X}, \text{axis}=1)$ 
% \State $\mathbf{X} \leftarrow \mathbf{X} - \boldsymbol{\gamma}_{min}$ 
% \State $\boldsymbol{\gamma}_{scale} \leftarrow \max(\mathbf{X}, \text{axis}=1)$
% \State $\mathbf{X} \leftarrow \mathbf{X} \oslash (\boldsymbol{\gamma}_{scale} + \epsilon)$ 
\State \textit{// Stage 1: Per-channel, page-Level Normalization}
\State $\boldsymbol{\gamma}_{min} \leftarrow \min(\mathbf{X}, \text{axis=1})$
\State $\boldsymbol{\gamma}_{scale} \leftarrow \max(\mathbf{X}, \text{axis=1}) - \boldsymbol{\gamma}_{min}$ 
\State $\mathbf{X} \leftarrow (\mathbf{X} - \boldsymbol{\gamma}_{min}) \oslash (\boldsymbol{\gamma}_{scale} + \epsilon)$

\vspace{0.1cm}
% \State \textit{// Stage 2: Chunk-wise Isolation}
% \State $\mathbf{X}_{view} \leftarrow \text{Reshape}(\mathbf{X}, [H \cdot (P/C), C])$ 
% \State $\boldsymbol{\mu} \leftarrow \text{Mean}(\mathbf{X}_{view}, \text{axis}=1)$
% \State $\mathbf{X}_{centered} \leftarrow \mathbf{X}_{view} - \boldsymbol{\mu}$
% \State $\boldsymbol{\sigma} \leftarrow \max(|\mathbf{X}_{centered}|, \text{axis}=1)$
% \State $\mathbf{X}_{final} \leftarrow \mathbf{X}_{centered} \oslash (\boldsymbol{\sigma} + \epsilon)$
\State \textit{// Stage 2: Per-channel, chunk-wise Isolation}
\State $\mathbf{X}_{view} \leftarrow \text{Reshape}(\mathbf{X}, [H \cdot (P/C), C])$ 
\State $\boldsymbol{\mu} \leftarrow \text{Mean}(\mathbf{X}_{view}, \text{axis=2}); \quad \mathbf{X}_{cent} \leftarrow \mathbf{X}_{view} - \boldsymbol{\mu}$
\State $\boldsymbol{\sigma} \leftarrow \max(|\mathbf{X}_{cent}|, \text{axis=2}); \quad \mathbf{X}_{final} \leftarrow \mathbf{X}_{cent} \oslash (\boldsymbol{\sigma} + \epsilon)$

\vspace{0.1cm}
\State \textit{// Stage 3: Non-Linear Transform}
\State $\mathbf{Y} \leftarrow \frac{\ln(1 + \alpha |\mathbf{X}_{final}|)}{\ln(1 + \alpha)} \cdot (2^7 - 1)$
\State $\mathbf{I} \leftarrow \text{Round}(\mathbf{Y})$ \Comment{Raw Integers}
\State $\mathbf{S} \leftarrow \text{Sign}(\mathbf{X}_{view})$

\vspace{0.1cm}
\State \textit{// Stage 4: Split-Stream Construction}
\State $\mathbf{I}_{bias} \leftarrow \mathbf{I} + 7$ \Comment{Round-Half-Down Bias}
\State $\mathbf{V}_{mag} \leftarrow \mathbf{I}_{bias} \gg 4$ \Comment{Extract Magnitude (Anchor)}
\State $\mathbf{V}_{recon} \leftarrow \mathbf{V}_{mag} \ll 4$ \Comment{Reconstruct Base}
\State $\mathbf{Q}_{res} \leftarrow \mathbf{V}_{recon} - \mathbf{I}$ \Comment{Calculate Correction Term}
\State $\mathbf{Q}_{anc} \leftarrow \mathbf{V}_{mag} \odot \mathbf{S} + \min(\mathbf{S}, 0)$ \Comment{Two's Complement Map} \label{line:sign_map}
\State \Return $\mathbf{Q}_{anc}, \mathbf{Q}_{res}, \{\boldsymbol{\gamma}_{min}, \boldsymbol{\gamma}_{scale}, \boldsymbol{\mu}, \boldsymbol{\sigma}\}$
\end{algorithmic}
\end{algorithm}

\begin{figure*}[t!]
  \centering

  % \hspace{-0.5cm}
  \subfigure[\sysname prefill instance.]{
    \begin{minipage}[t]{0.31\linewidth}
      \vspace{-0.00in}
      \begin{center}
        \includegraphics[width=\textwidth]{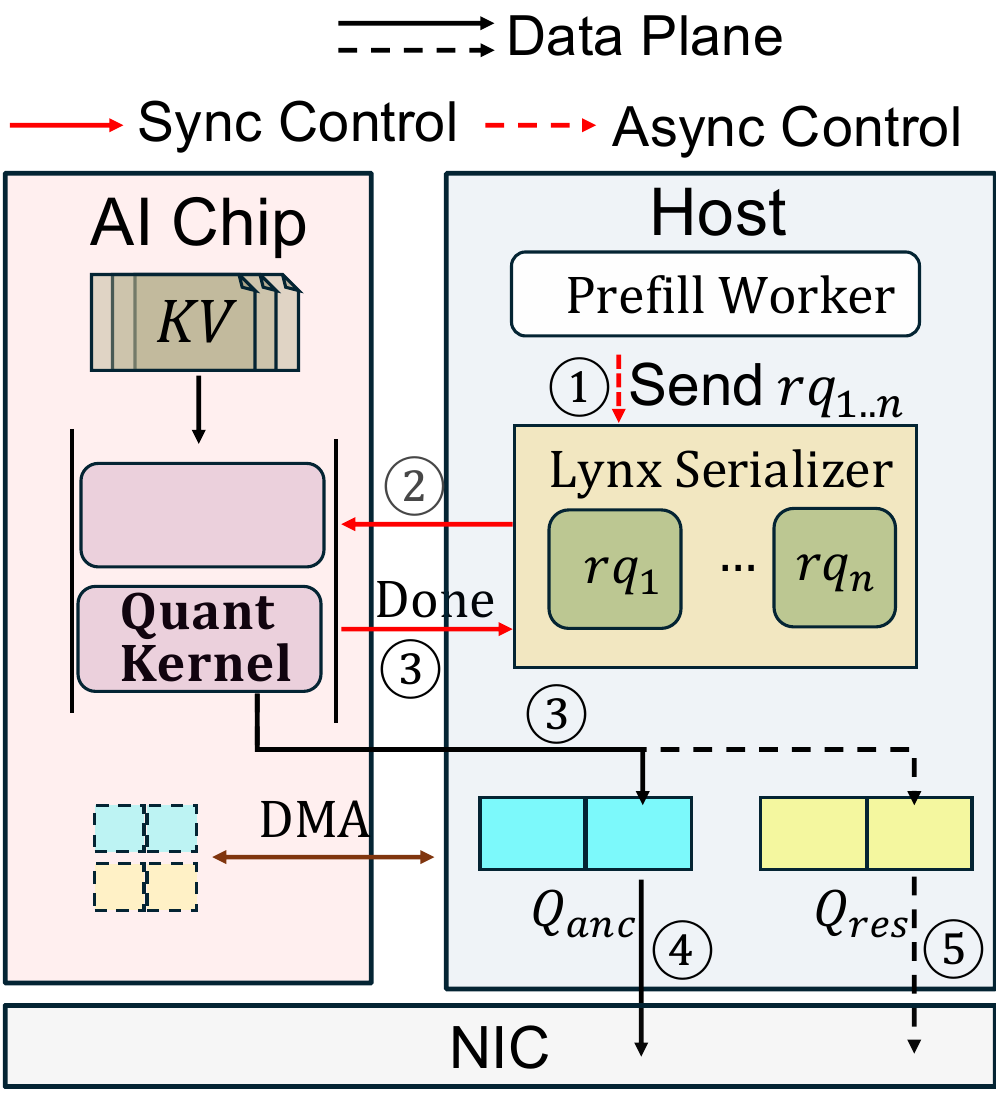}
      \end{center}
    \end{minipage}
    \label{subfig:overview-matrix-a}
  }
  % \vspace{2mm}
  \hfill
%
  % \hspace{-0.2cm}
  \subfigure[Anchor Phase: Receiving $Q_{anc}$ on \sysname \mbox{decode} instance.]{
    \begin{minipage}[t]{0.31\linewidth}
      \vspace{-0.00in}
      \begin{center}
        \includegraphics[width=\textwidth]{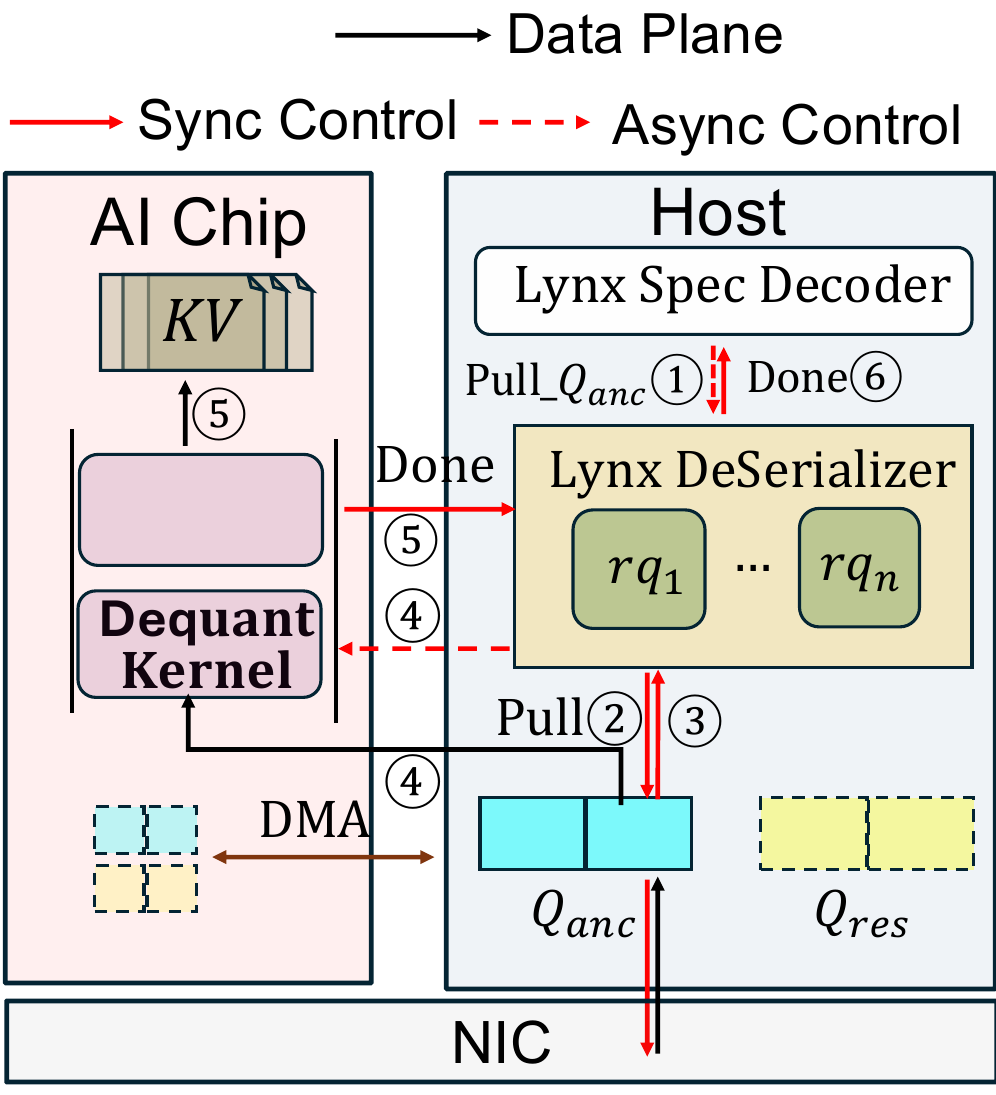}
      \end{center}
    \end{minipage}
    \label{subfig:overview-matrix-c}
  }
  \hfill
  % \hspace{-0.3cm}
  \subfigure[Residual Phase: Speculative decoding in parallel with receiving $Q_{res}$ on the decode instance.]{
    \begin{minipage}[t]{0.30\linewidth}
      \vspace{-0.01in}
      \begin{center}
        \includegraphics[width=\textwidth]{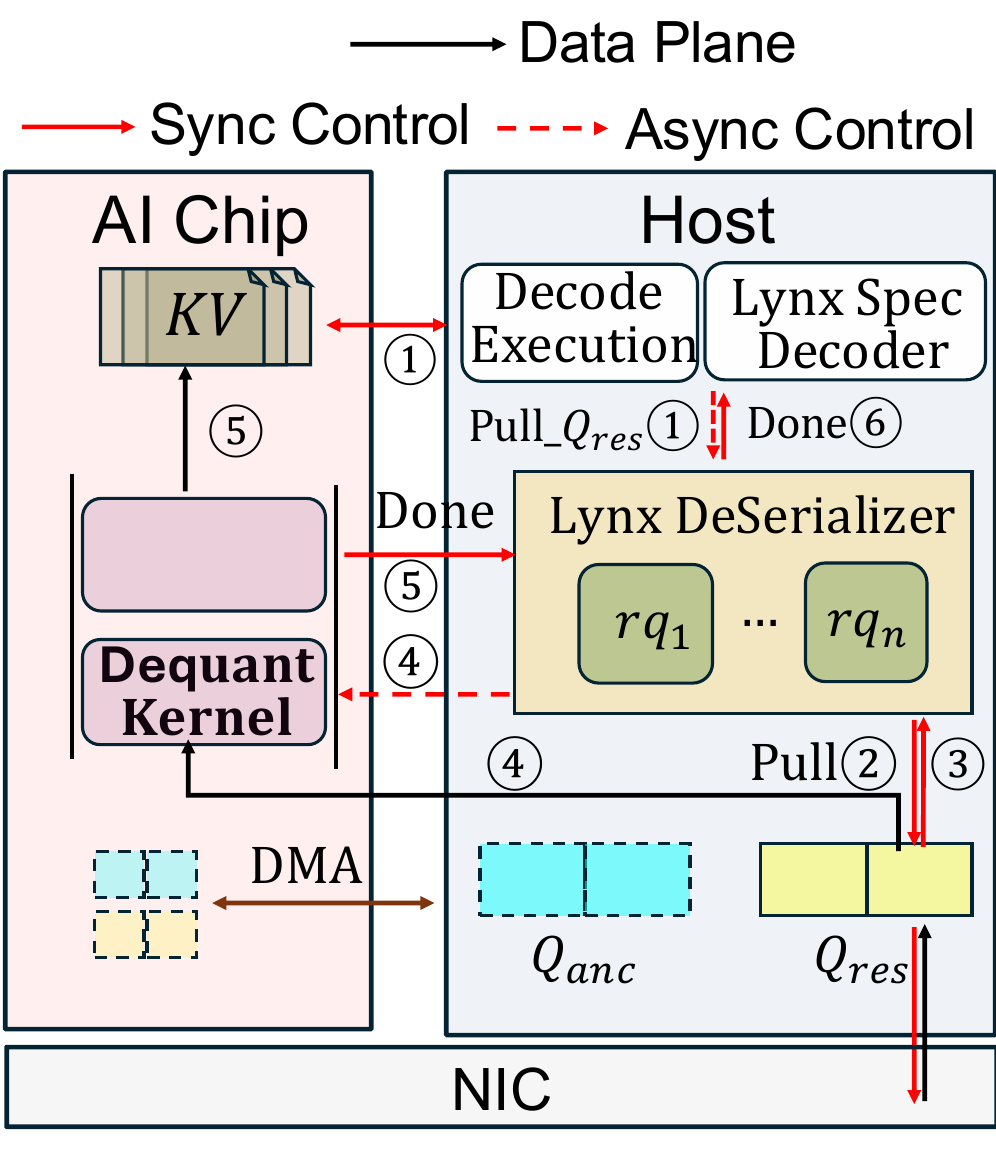}
      \end{center}
    \end{minipage}
    \label{subfig:overview-matrix-d}
  }

  \vspace{-4mm}

  \caption{The \sysname's prefill and decode workflows described in \S\ref{sec:bit_stream_pipe}, where the red arrows refer to synchronous control operations, the dashed red arrows refer to asynchronous control operations and the black arrows refer to the computation on the AI chip and the KV cache data movement. $rq_1, \cdots, rq_n$ are pieces of control messages, each of which encodes a request to serialize and send, or fetch and deserialize a page of KV cache data. }%components and interactions. }
  \label{fig:workflow-matrix}
  % \vspace{-0.2cm}
\end{figure*}

The complete procedure is formalized in Algorithm~\ref{alg:quantization} and consists of four stages. 
\\\textbf{1) Page-Level Normalization.} We process the cache in logical pages (e.g. $P=256$ tokens) per channel, complying with the spread out paged attention employed by inference engines~\cite{kwon2023efficient}. First, we normalize the floating-point distribution into a positive bounded range. We calculate the global minimum ($\gamma_{min}$) and a scaling factor ($\gamma_{scale}$) for the page, mapping the values to a normalized domain $[0, 1]$. 
\\\textbf{2) Per-channel, chunk-wise Outlier Isolation.} We further subdivide the page into local chunks to isolate outliers in a per-channel, per-chunk manner. A chunk refers to a consecutive sequence of entries in a KV cache that reside within a channel and span tokenwise across a consecutive number of tokens (\eg, $C=32$). The design choice of per-channel outlier isolation is motivated by the fact that extreme outliers usually concentrate within a few channels and the entries within each channel are of the same order of magnitude (\Cref{fig:kv-cache-distribution}). We further partition entries in each channel into chunks, as a lower chunk size further reduces the quantization range, resulting in a lower compression error. 

Specifically, for each chunk, we calculate a local mean $\mu_{chunk}$ and a local scalar $\sigma_{chunk}$. These parameters are used to center the distribution around zero (by subtracting the mean) and absorb the magnitude of any local outliers (by dividing by the scalar). The use of a local scalar ensures that an extreme value in one chunk only inflates the scaling factor $\sigma_{chunk}$ of the local region, rather than dictating the quantization step size for the entire token page.
\\\textbf{3) Non-Linear Transformation.} We apply a logarithmic transformation ${ln}$ to the normalized data using the standard $\alpha$-law formulation, where $\alpha$ controls the skewness:
\begin{equation}
y = \frac{\ln(1 + \alpha |\mathbf{x}_{norm}|)}{\ln(1 + \alpha)} \cdot (2^7 - 1), \alpha > 0
\end{equation}
The inverse of this transformation implies that the reconstructed value $\hat{x}$ follows an exponential curve relative to the stored absolute value of integer $y$:
\begin{equation}
    |\hat{x}| \propto (1 + \alpha)^{\frac{y}{127}}
\end{equation}
\\\textbf{4) Split-Stream Construction.} We construct the integer bits represent a functional equivalence to floating-point:
\begin{itemize}[leftmargin=*]
    \item \textbf{Anchor Stream (MSBs $\cong$ Exponent):} The top 4 bits partition the dynamic range into geometric intervals. For a quantized anchor value $q_{anc} \in \mathbf{Q}_{anc}$, the reconstructed magnitude scales with $(1+\alpha)^{q_{anc}}$. This effective exponentiation allows the Anchor Stream alone to capture the Order of Magnitude required to calculate attention scores during speculative decoding.
    \item \textbf{Residual Stream (LSBs $\cong$ Mantissa):} The bottom 4 bits perform linear interpolation within the geometric interval defined by the anchor. To minimize quantization error, we employ a Round-Half-Down bias (+7) during extraction, ensuring the residual encodes a centered, signed correction term for maximum precision.
\end{itemize}
Crucially, we employ a \textit{sign-aware mapping} (Algorithm~\ref{alg:quantization}, line~\ref{line:sign_map}) to map $-0$ to distinct negative integers. This preserves the sign bit even for small-magnitude values, which is vital for maintaining the correct sign of attention scores in the low-precision regime.

\subsection{Prioritized Transmission Mechanism}
\label{sec:bit_stream_pipe}
%\gingfung{ideally, we really should reference the figures for these components and steps.}

%Atop our hierarchical quantization algorithm, we are ready to elaborate \sysname's main architecture (\Cref{fig:workflow-matrix}).
%Having established the quantization format, we now detail \sysname's system architecture (\Cref{fig:workflow-matrix}) that executes the prioritized transfer and speculative generation (Stages 2 and 3). 
\sysname implements a non-blocking architecture composed of two logical modules (\Cref{fig:sys-overview}): a Stream SerDes that manages data handling and transmission, and a Speculative Decoder that orchestrates the generation workflow. The serializer and deserializer manage the quantization and transmission of paged KV cache objects in a pipelined fashion. 

%Overall, our architecture includes a serializer and a deserializer that implement the pipelined priority streaming protocol and serializes the quantized objects efficiently, and a non-blocking \sysname speculative decoder as the serving runtime that controls the workflow of speculative generation with overlapped KV transfer~\Cref{fig:sys-overview}.

\subp{Prefill Instance Workflow (\Cref{subfig:overview-matrix-a}).} The process initiates when the Prefill Runner submits requests ($rq_{1..n}$) to the serializer \ding{172}. The Serializer dispatches these to the quantization kernel \ding{173}, which processes each KV object $\mathbf{X}$ to generate the quantized components ($Q_{anc}, Q_{res}$), and metadata ($\gamma, \mu, \sigma$).
To maximize DMA throughput, the kernel enforces a contiguous memory layout: writing the compacted $Q_{anc}$ and metadata to a transmission buffer via DMA \ding{174}. Finally, the serializer pushes the $Q_{anc}$ \ding{175} and $Q_{res}$ \ding{176} to the NIC via two separate queues.

Crucially, \sysname implements pipeline overlapping, the network transmission of object $i$ runs concurrently with the quantization of object $i+1$, effectively hiding the serialization overhead, ensuring the network link remains saturated.

\subp{Decode Instance Workflow.} On the decode instance side, the deserializer reconstructs the cache in two phases:
\begin{itemize}[leftmargin=*]
    \item \textbf{Anchor Phase (\Cref{subfig:overview-matrix-c}):} Token generation commences when the Anchor state is available. The Speculative Decoder issues a \texttt{Pull\_$Q_{anc}$} command \ding{172}, triggering the deserializer to fetch incoming $Q_{anc}$ data from the NIC \ding{173}.
    \ding{174} Upon receiving the $Q_{anc}$, the deserializer then invoke the dequantization kernel with $Q_{anc}$ and the metadata \ding{175}, finally scattered into the KV pages \ding{176}.
    Upon completion \ding{177}, the model immediately accesses the Anchor state for execution, while metadata is cached for the next phase.
    \item \textbf{Residual Phase (\Cref{subfig:overview-matrix-d}):} While the decoder executes speculatively, it concurrently issues a \texttt{Pull\_$Q_{res}$} command \ding{172}.
    The deserializer retrieves the buffered $Q_{res}$ stream \ding{173}-\ding{174}, combines it with the stored metadata, and invoke the dequantization kernel \ding{175}. The kernel refines the KV cache to full precision ($Q_{full}$) \ding{176}, allowing the decoder to transition the verification stage upon the done signal \ding{177}.
\end{itemize}

\iffalse
\begin{itemize}[leftmargin=*]
    \item \textbf{Anchor Phase (\Cref{subfig:overview-matrix-c}):} \ding{172} Upon $Q_{anc}$ arrival, \ding{173} $Q_{anc}$ is immediately pull into the dequantization kernel via the deserializer \ding{174}-\ding{175}, and scattered into the KV pages \ding{176}-\ding{177}. This allows the model to immediately access the Anchor state and ready for execution. While the metadata is cached to facilitate the later arrival of $Q_{res}$.
    \item \textbf{Residual Phase (\Cref{subfig:overview-matrix-d}):} \ding{172} Once the residual stream arrives, \ding{173}-\ding{174} the deserializer combines the $Q_{res}$ with the stored metadata, and \ding{175} invoke the dequantization kernel to refine the KV cache to full precision $Q_{full}$. \ding{176} Finally, scattered into the KV pages and \ding{177} move onto the next verification stage.
\end{itemize}
\fi
In our prototype, Anchor buffers are always drained before Residual buffers, enforcing strict priority at the stream level. Unlike layer-wise pipelining, which still blocks on per-layer completeness. %\sysname exposes partial KV state within a layer, enabling decoding to begin earlier. 

\subp{Speculative Decoder.} The decoder serves as the central serving routine, managing the SerDes operations as asynchronous background coroutines. 
Token generation triggers immediately upon the completion of the Anchor stream deserialization (Step \ding{177} in \Cref{subfig:overview-matrix-c}). During this interval, the model operates on the partial state of the KV cache, marking output tokens as speculative drafts ($s_1, \cdots, s_t$).

Before generating each subsequent token, the decoder checks the background Residual transfer status. Once the Residual stream is fully received and dequantized, the decoder transitions from generation to verification, where the draft sequence $s_1, \dots, s_t$ is validated against the now-available high-precision KV cache $Q_{full}$ (detailed in \Cref{sec:verification}).

\subsection{Verification and Correction}
\label{sec:verification}

%The verification and correction of \sysname is a lightweight process taking place once the deserialization of the residual stream completes and the KV cache is updated into $8$-bit precision. The process involves a one-time forward pass of the model against the speculative sequence $s_1, \cdots, s_t$, deciding on the longest prefix of the speculative sequence that can be accepted. 
%\sysname adapts the verification algorithm of speculative decoding proposed in~\cite{leviathan2023fast}. Here the draft model corresponds to the LLM parameterized by the MSBs of the KV cache $Q_{anc}$ and the target model is the LLM parameterized with the high INT8 precision KV cache $Q_{full}$. 
Upon the completion of the Residual transfer, \sysname transitions to the verification phase to ensure valid generation quality. We adapt the speculative decoding framework~\cite{leviathan2023fast,chen2023accelerating} to the context of progressive KV transfer and quantization. 

%In our setup, we instantiate the generic draft model $q(\cdot)$ from Equation~\ref{eqn:spec_dec} as the LLM operating on the Anchor-only KV cache ($Q_{anc}$), while the target model $p(\cdot)$ corresponds to the same LLM utilizing the restored high-precision cache ($Q_{full}$).
%Specifically, during the speculative generation with $Q_{anc}$, we generate the speculative tokens $s_1, \cdots, s_t$ as well as the probabilistic distribution vectors of $p_1(x), \cdots, p_t(x)$, where $p_i(x)$ denotes the probability that $x$ is sampled at the $i$th token. During the speculative verification, we feed the speculative tokens $s_1, \cdots, s_t$ into the target model with $Q_{full}$ KV cache to generate the verification tokens $v_1, \cdots, v_{t+1}$ along with the probabilistic distribution vectors $q_1(x), \cdots, q_{t+1}(x)$. Here $o_i$ represent that the target model with $Q_{full}$ samples $v_i$ on the condition that $s_1, \cdots, s_{i-1}$ are the prefix tokens. Finally  With $\{s\}, \{p\}, \{v\}$ and $\{q\}$, we accept token $s_i$ with probability $acc_i=\frac{p_i(v_i)}{q_i(v_i)}$, and we calculate the longest prefix $l \le t$ of the accepted tokens to be converted as the verified output tokens $o_1, \cdots, o_l=s_1, \cdots, s_l$.
The verification involves a single parallel forward pass using $Q_{full}$ to validate the speculative sequence $s_1, \dots, s_t$. %By comparing the output distributions of $Q_{anc}$ against those from $Q_{full}$ via Equation~\ref{eqn:spec_dec}, 
\sysname identifies the longest prefix of tokens, accepting valid tokens and correcting the first divergence without re-computation.
%The aforementioned algorithm guarantees that the distribution of the resulting output sequence follows the same distribution as those produced by generating solely with INT8-precision KV cache $Q_{full}$, thereby ensuring \sysname's INT8-equivalent inference accuracy. A modified version of the algorithm is further raise the acceptance probability $acc_i$ to have more speculative tokens to be accepted, but this comes at the tradeoff of compromised inference accuracy. In~\Cref{fig:spec-acceptance} in~\Cref{sec:eval}, we show that the acceptance rate of the speculative tokens is sufficiently high, and thus we follow the original algorithm to guarantee no inference accuracy degradation.

This verification protocol guarantees that the resulting output sequence follows the distribution produced by the target model ($Q_{full}$), thereby ensuring that \sysname maintains full-precision inference accuracy. While relaxed verification schemes exist that trade accuracy for higher acceptance rates, we find them unnecessary in our context. As shown in \Cref{fig:spec-acceptance} (Section~\ref{sec:eval}), the high fidelity of the Anchor stream already yields a sufficiently high acceptance rate, allowing us to retain the guarantee of the original algorithm without performance penalty.

\section{Implementation}
We have implemented our \sysname prototype in ${\approx}2k$ LoC of Ascend-C kernels~\cite{ascend} code for Ascend NPUs, and ${\approx}2k$ LoC in Python, that is integrated into vLLM-Ascend~\cite{vllmascend}, with LMCache-Ascend~\cite{lmcascend} as the KV transfer connector.

The Python code implements \sysname's progressive execution to achieve computation-communication overlaps and speculative verification to ensure the same accuracy as decoding with the higher precision (i.e., INT8) KV cache. The Ascend-C kernels implement the core logic of \sysname's KV compression and decompression. 

\subp{Compression Kernel.} Similar to GPUs, NPU-based compression is a memory-bound task limited primarily by DRAM bandwidth. To support efficient two-phase transfer, our kernel must serialize data such that each stream (Anchor and Residual) forms a contiguous block in physical memory.
We implement this by computing dynamic memory offsets for the bit-packed payload: the kernel extracts the MSBs from input scalars, \ie, packing two 4-bit segments into a single uint8 byte, and writes them sequentially to the Anchor buffer, followed immediately by the shared scalars ($\gamma, \sigma$). To hide memory latency, we utilize the Ascend NPU's AI Vector core and Unified Buffer (UB) to implement a Double-Buffering pipeline (Ping-Pong)~\cite{ascendc_guide,LIANG202075}. As illustrated in Algorithm~\ref{alg:quantization}, the kernel loads a chunk of KV data into the Ping buffer while simultaneously compressing and writing out the Pong buffer, ensuring the execution units remain saturated.

\section{Evaluation}\label{sec:eval}

This section presents the end-to-end testbed evaluation of \sysname. The evaluation includes popular LLM inference workloads with mid-size models and compares \sysname with a range of uncompressed and compressed baselines. 

Across three models and three long-context workloads, \sysname is the only approach that simultaneously achieves 1) BF16-equivalent accuracy and 2) INT4-level Time to First Token. Both gains hold when scaling \sysname to up to $128$K context length and $50$Gbps bandwidth. In contrast, CacheGen and INT4-compression sacrifice inference accuracy, while INT8-compression leads to both higher KV transfer latency and slightly larger compression error than \sysname with $Q_{full}$ KV cache (\sysname-INT8).% that does not result in observable end-to-end accuracy loss.

%Results show that, compared to the baselines, \sysname is the only solution that achieves BF16-equivalent inference accuracy, while reaching INT4-level KV transfer delay and inference latency. 

\subsection{Evaluation setups}\label{subsec:setups}

\subp{Testbed.} The experiments are conducted on a testbed comprising two Atlas A2 servers~\cite{huawei_atlas800ia2}. Each server is equipped with $8$ high-performance 910B4 NPUs, each of which has a $32$GB on-device HBM memory. %The two servers are interconnected with a $200$Gbps RDMA NIC, and every pair of NPUs are connected via HCCS with unidirectional $224$Gbps.%\Wenchen{TO CHECK.}

\subp{Inference deployment setups.} We evaluate \sysname on a disaggregated prefill-decode setup with two vLLM instances (one per server) interconnected by our testbed network. To emulate the network bottlenecks characteristic of long context scenarios (e.g., $>1$M tokens) within our testbed's memory constraints, we implement a rate-limiter on the KV connector. We restrict inter-server bandwidth from $10$Gbps to $50$Gbps as this effectively recreates the high transfer-to-compute latency ratios typical of very long context workloads running on standard high-speed cluster networks\footnote{Transferring the KV cache for a 128K tokens over a 10Gbps link $\approx$ same latency ratio as transferring a 1M tokens over an 80Gbps link}.
%. Since scaling tokens length increases memory volumne linearly, our lower-bandwidth setup effectively simulates the transfer bottlenecks of future million-tokens workloads.}.

%The instances are interconnected by our testbed network. On the prefill server, we implemented a rate limitation mechanism for our KV connector to emulate different bandwidth settings available between the prefill and the decode server. Unless otherwise mentioned, the bandwidth is restricted to $10$Gbps\marco{CRITIC}. Focusing on long-context inference, we constructed several system prompts of different lengths that are shared across different requests. The system prompts include detailed instructions of how to answer users' questions followed optionally by several few-shot examples, both of which improve LLMs' inference accuracy~\cite{}. The KV cache data of the shared system prompts is computed before processing the main inference workloads, so that the shared prompts' KV cache is reused during inference, and the overhead of recomputing the KV cache is eliminated.

\subp{Models.} We select three popular LLMs of different model families: namely LLaMA 3.1 8B Instruct~\cite{grattafiori2024llama}, Qwen 3 32B~\cite{qwen3} and Mistral 3 24B Instruct~\cite{mistral24b}. Both LLaMA 8B and Mistral 24B support a context window up to $128$K tokens. Qwen 32B instead supports a context length of $32$K tokens. To handle requests longer than $32$K tokens, we adopt YaRN method~\cite{peng2024yarn} as a RoPE scaling technique to support contexts of $128$K tokens.

\begin{table}[h!]
    \centering
    \resizebox{1\linewidth}{!}{
    \begin{tabular}{|c|c|c|c|c|} \hline
        Dataset & Task & Nsamples & Length & Metric \\ \hline
        \multirow{2}{*}{MMLU-Pro} & Few-shot & \multirow{2}{*}{512} & \multirow{2}{*}{16/32/64/128K} & \multirow{2}{*}{Accuracy} \\
                                 & CoT QA   &  & &  \\ \hline
        Needle & Retrieval & 512 & 10K & Rouge-L  \\ \hline
        QMSum & Summarization & 200 & 10K & Rouge-L \\ \hline
    \end{tabular}
    }
    \caption{LLM inference evaluation datasets and target metrics (Section~\S\ref{subsec:setups}).} %In MMLU-Pro with few-shot prompting, several input-output examples are provided in the shared system prompt to guide LLM's inference performance, and we control the context length by altering the number of examples given in few-shot prompting.}
    \label{tab:datasets}
\end{table}

%gingfung don't think we need them +1
%We constructed a suite of system prompts with varying lengths shared across requests. These prompts contain detailed instruction to answer users' questions and optional few-shot examples, which are known to enhance inference accuracy~\cite{wei22cotp,rubinetal2022learning}. To isolate the performance of the transfer mechanism, the KV cache for these shared prompts is pre-computed and resident in memory, eliminating re-computation overhead during the experimental runs. 
\subp{Datasets.} \sysname is evaluated against three inference tasks with different datasets, as summarized in~\ref{tab:datasets}. %\marco{Can we list all of the tasks here and then describe?} 
In MMLU-Pro~\cite{wang2024mmlu}, we run multiple-choice question answering with chain-of-thoughts (CoT) generation~\cite{wei22cotp,rubinetal2022learning}. We modulate the shared context length (defaulting to $16$K in~\Cref{subsec:e2e-accuracy} and~\ref{subsec:e2e-latency}) by varying the number of few-shot examples in the system prompt.
%where we include a varying number of few-shot examples in the system prompt to control variable context length (defaulted to $16$K in~\Cref{subsec:e2e-accuracy} and \Cref{subsec:e2e-latency}). 
%The model generates a response with CoT tokens followed by the final choice as the answer. The benchmark score measures the accuracy of the model's answers.
We report the multiple-choice answer accuracy as the target metric.
In the multilingual Needle-in-the-haystack dataset~\cite{NEURIPS2024_24a8968a}, the task is to retrieve specific information as appeared in the context. The average context length is ${\approx}10k$ tokens. In QMSum~\cite{zhongetal2021qmsum}, the target task is to generate a short summary to questions given by users from a long report. For both Needle and QMSUM, Rouge-L~\cite{lin2004automatic} is applied to measure the similarity between the generated response of the model and the reference response. Throughout the paper we will use the general term ``\textit{inference accuracy}" to refer to the respective evaluation metrics of different datasets.

\subp{Baselines.} We compared \sysname against 4 KV-transfer baselines atop vLLM-Ascend as the inference engine: i) the uncompressed baseline that transfers KV data directly in BF16~\cite{kalamkar19bf16}; ii) the INT8 and iii) INT4 uniform quantization schemes, which uniformly quantize KV cache into INT8 and INT4 formats respectively for KV transfer; and iv) CacheGen~\cite{liu2024cachegen}'s quantization strategy, which includes several optimization techniques such as delta encoding and token-wise compression. Following CacheGen's original setup, we configured it to use a mixed INT4 and INT8 precision for quantization.

Due to the lack of an official CacheGen implementation for Ascend NPUs, we developed a best-effort port to the hardware architecture. While this implementation is not fully optimized, and may exhibit higher computational overhead, it serves as a functional baseline to evaluate CacheGen's compression logic. Thus, we only report CacheGen's inference accuracy, demonstrating its inaccurate compression introduces considerable accuracy degradation, which is independent of hardware performance.

\begin{figure*}
    \centering
    \begin{minipage}[t]{0.4\linewidth}{
		\vspace{-0.00in}
		\begin{center}
		\includegraphics[width=\textwidth, ]{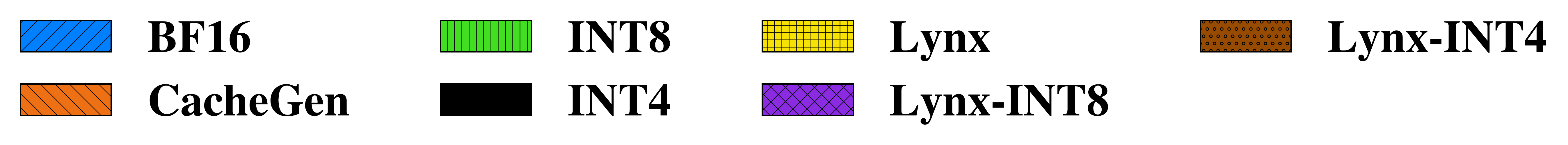}
		\end{center}
		}
        \end{minipage}

    % \hspace{-0.2cm}
        \subfigure[MMLU LLaMA]{
		\begin{minipage}[t]{0.155\linewidth}{
		\vspace{-0.00in}
		\begin{center}
		\includegraphics[width=\textwidth, ]{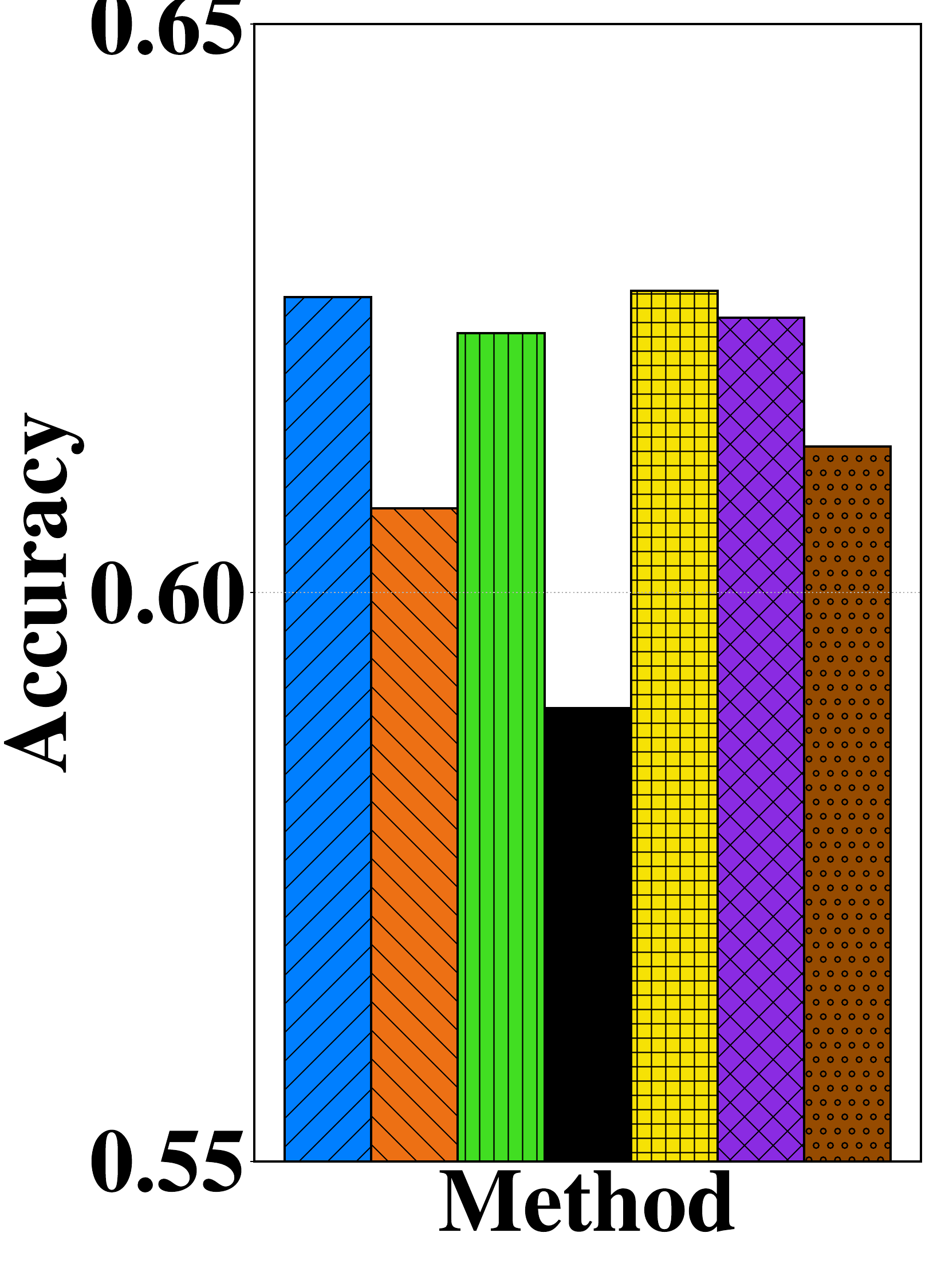}
		\end{center}
            \vspace{-0.1cm}
		}
		\label{subfig:accuracy-mmlu-llama}
		\end{minipage}
	}
    \hspace{-0.2cm}
        \subfigure[MMLU Qwen]{
		\begin{minipage}[t]{0.155\linewidth}{
		\vspace{-0.00in}
		\begin{center}
		\includegraphics[width=\textwidth, ]{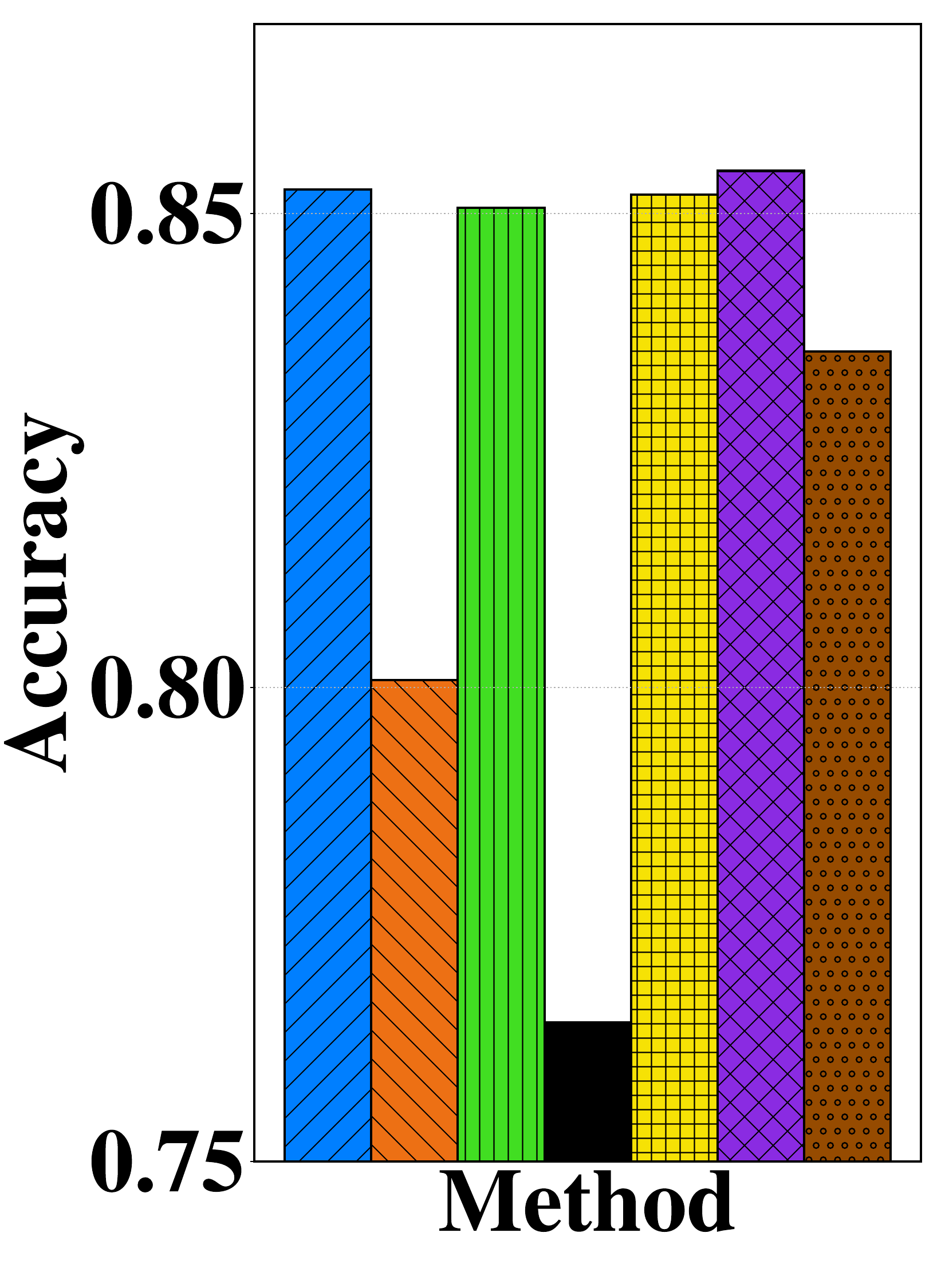}
		\end{center}
        \vspace{-0.1cm}
		}
		\label{subfig:accuracy-mmlu-qwen}
		\end{minipage}
	}
    \hspace{-0.2cm}
        \subfigure[Needle LLaMA]{
		\begin{minipage}[t]{0.155\linewidth}{
		\vspace{-0.00in}
		\begin{center}
		\includegraphics[width=\textwidth, ]{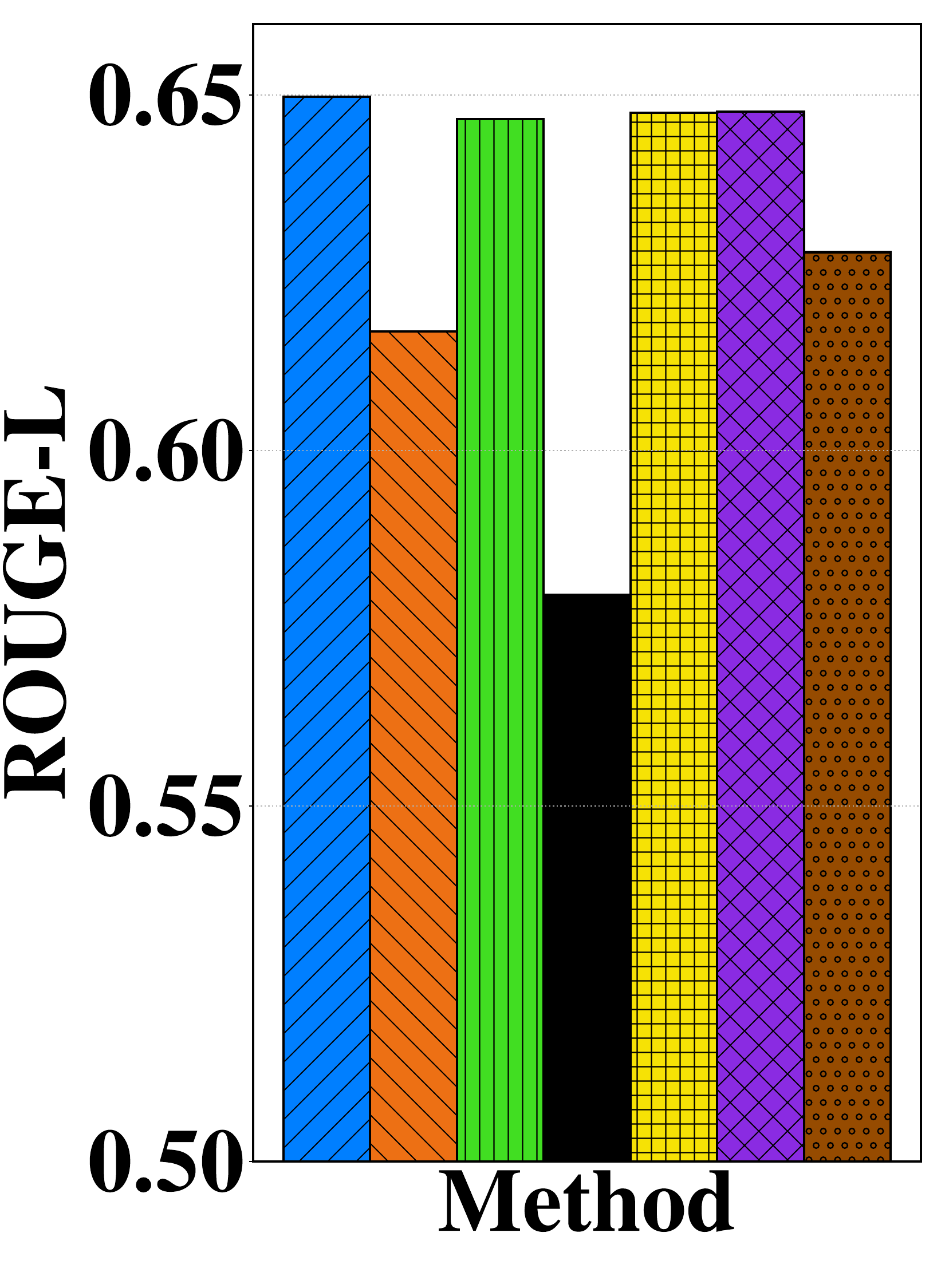}
		\end{center}
        \vspace{-0.1cm}
		}
		\label{subfig:accuracy-needle-llama}
		\end{minipage}
	}
    \hspace{-0.2cm}
        \subfigure[Needle Mistral]{
		\begin{minipage}[t]{0.155\linewidth}{
		\vspace{-0.00in}
		\begin{center}
		\includegraphics[width=\textwidth, ]{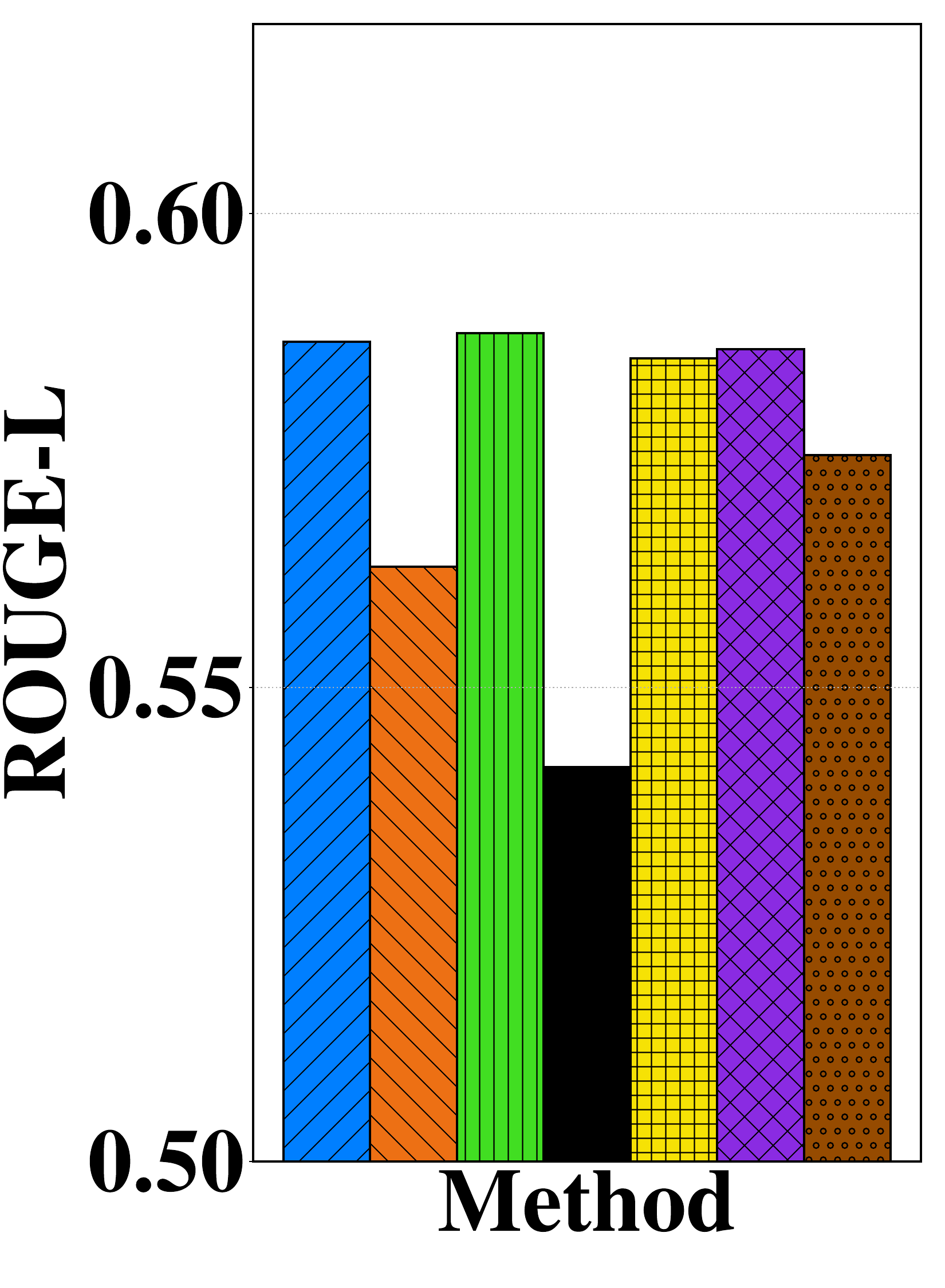}
		\end{center}
        \vspace{-0.1cm}
		}
		\label{subfig:accuracy-mistral-needle}
		\end{minipage}
	}
    \hspace{-0.2cm}
        \subfigure[QMsum Qwen]{
		\begin{minipage}[t]{0.155\linewidth}{
		\vspace{-0.00in}
		\begin{center}
		\includegraphics[width=\textwidth, ]{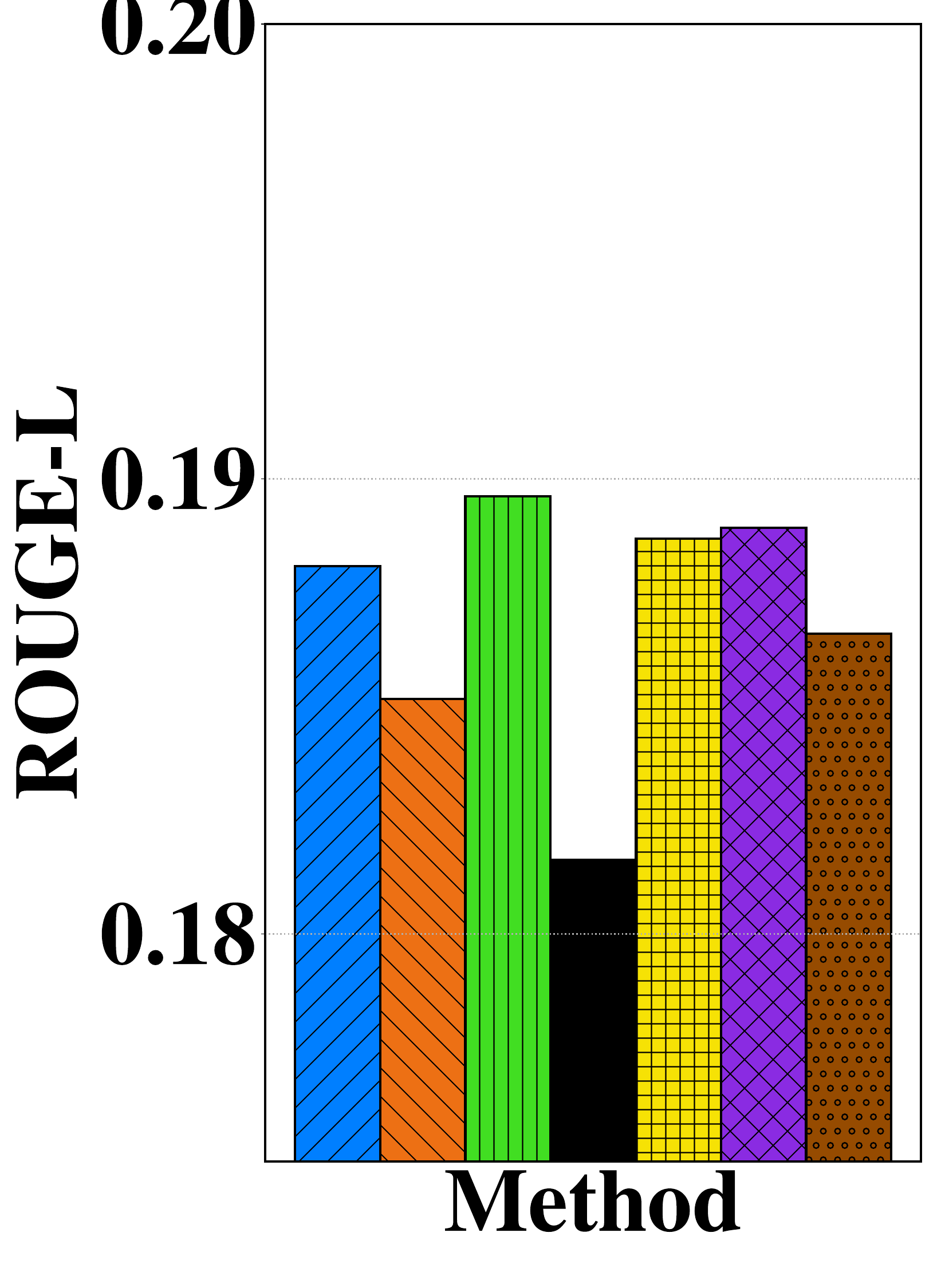}
		\end{center}
        \vspace{-0.1cm}
		}
		\label{subfig:accuracy-mistral-qwen}
		\end{minipage}
	}
    \hspace{-0.2cm}
        \subfigure[QMsum Mistral]{
		\begin{minipage}[t]{0.155\linewidth}{
		\vspace{-0.00in}
		\begin{center}
		\includegraphics[width=\textwidth, ]{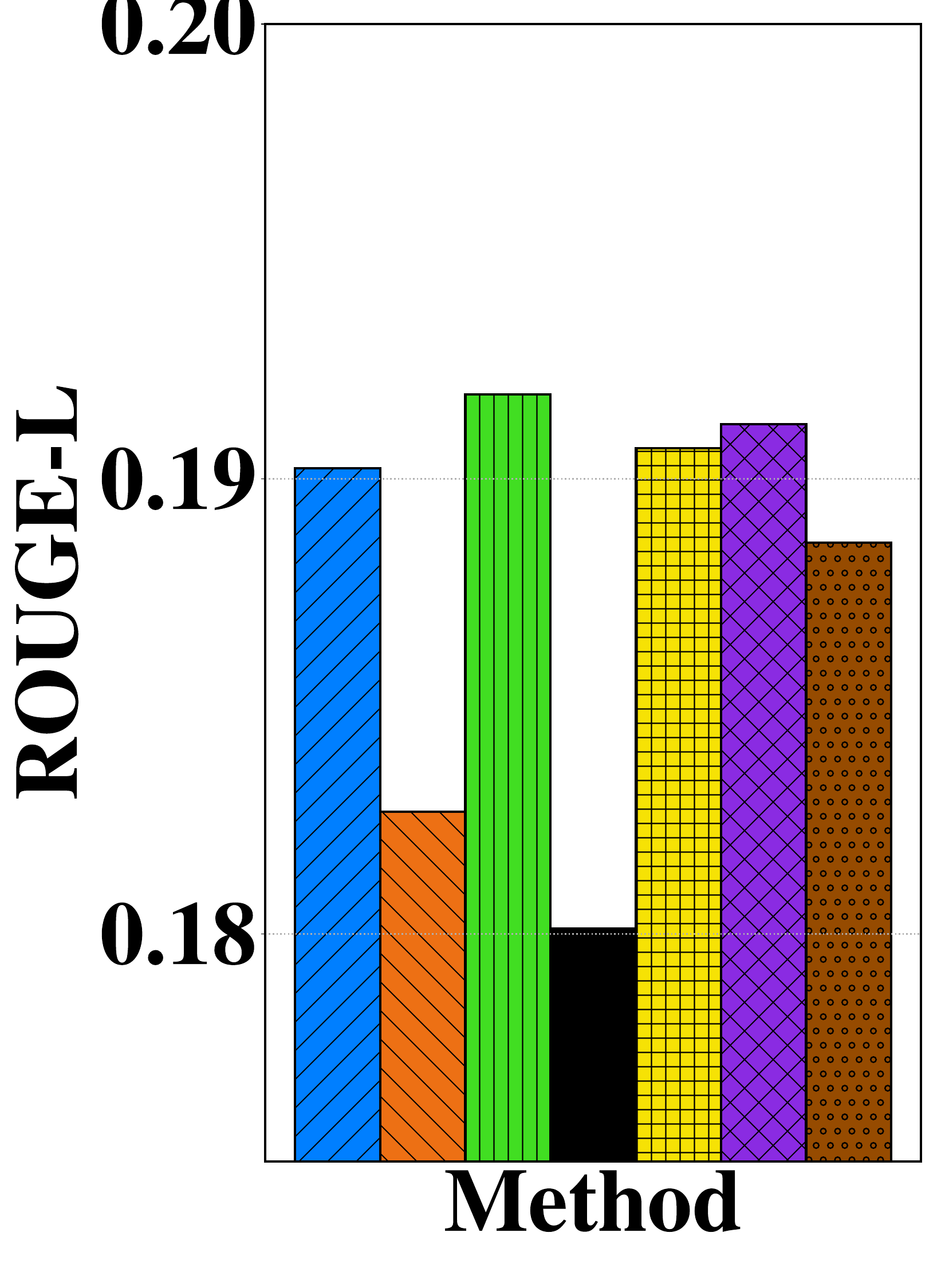}
		\end{center}
        \vspace{-0.1cm}
		}
		\label{subfig:accuracy-mistral-qmsum}
		\end{minipage}
	}

    \vspace{-0.3cm}
    \caption{End-to-end inference accuracy across different LLM inference workloads. \sysname hierarchical quantization schemes achieves similar accuracy across datasets when compared to both BF16 and Int8.}%The metric for MMLU is set as the accuracy for multiple-choice questions and the metric for Needle and QMsum is set as ROUGE-L which measures the similarity between the model's output sequence and a reference answer sequence. We attribute the negligible accuracy differences among INT8, \sysname, \sysname-INT8, and BF16 to random variation of the random generation process and \sysname's random verification.}
    \label{fig:accuracy-e2e}
    \vspace{-0.0cm}
\end{figure*}

\subp{\sysname's configurations.} We implemented three versions of \sysname prototypes, namely \sysname, \sysname-INT4 and \sysname-INT8. \sysname-INT4 and \sysname-INT8 only quantize KV cache into a single precision using our proposed hierarchical quantization algorithm (\Cref{sec:hierarchical_quant}), without split-streaming KV transfer for overlapping KV transfer with decoding. In \sysname, we set the chunk size $C=32$ and the page size to $256$.%, applying a per-page INT8 quantization and per-chunk INT4 quantization. 

We capped the number of speculative tokens generated by \sysname to $64$, as we observe the acceptance rate diminishes for speculative tokens beyond $64$ (as shown in \Cref{fig:spec-acceptance}). After generating $64$ speculative tokens, \sysname waits for KV transfer completion for the LSBs portion if not finished yet.

\subp{System metrics.} We measured both the inference accuracy (as defined above) and latency throughout experiments. To quantify the latency benefits of progressive execution, we introduce the Time-to-$k$th-Token (TTKT) metric, defined as the cumulative latency from the start of decoding until the generation of the $k$th token. For \sysname, the metric only accounts for valid tokens, i.e., \textit{accepted} speculative tokens or tokens generated after \sysname's residual stream transfer and verification of speculative tokens.

\subsection{End-to-end inference accuracy}\label{subsec:e2e-accuracy}

We first elaborate \sysname's end-to-end inference accuracy compared with the baselines, and explain our end-to-end accuracy results with their respective compression error that reflects the accuracy degradation of each approach.  % We show that \sysname achieves high accuracy that is on par with the uncompressed BF16 baseline across all our LLM inference workloads, while CacheGen and INT4 exhibit significant accuracy degradation.

\subp{End-to-end accuracy.} Figure~\ref{fig:accuracy-e2e} shows the inference accuracy results of the three \sysname implementations and the baselines. The accuracy of INT8, \sysname, and \sysname-INT8 are of little statistical difference (\ie, within the $\pm 0.3\%$ range) from BF16's accuracy that can be regarded as random variation. This suggests that both INT8 and \sysname-INT8 are accurate enough, and LLMs can typically tolerate the minimal compression error incurred.

%\marco{would move the following into design section, and here only recall} \gingfung{agreed}
%We recall that \sysname's split stream transfer and our verification mechanism (as presented in~\Cref{sec:design}) guarantees that the resulting output distribution is the same as decoding entirely with INT8 KV cache. This ensures the same accuracy in expectation between \sysname and \sysname-INT8.

In contrast, we observe that all of INT4, \sysname-INT4 and CacheGen's compression algorithms incur sufficiently high accuracy degradation across our inference workloads. For example, INT4, \sysname-INT4 and CacheGen bring accuracy drops of $8.7\%$, $1.7\%$ and $5.1\%$ respectively on MMLU + Qwen, and ROUGE-L drops of $4.5\%$, $0.9\%$, and $2.7\%$ respectively on Needle + Mistral. Furthermore, \sysname-INT4 achieves higher accuracy than CacheGen and INT4, thanks to the optimization techniques of \sysname's quantization algorithm (\Cref{sec:hierarchical_quant}).

\begin{table}[h!]
    \centering
    \resizebox{1\linewidth}{!}{
    \begin{tabular}{|c||c|c|c|c|c|} \hline
        Method & CacheGen & INT4 & INT8 & \sysname & \sysname-INT4 \\ \hline
        vNMSE & 0.11 & 0.53 & 0.0042 & 0.00017 & 0.015 \\ \hline
    \end{tabular}
    }
    \caption{vNMSE for the LLaMA 8B + MMLU workload. \sysname and its variants, outperforming standard INT8, INT4 and CacheGen quantization schemes.}
    \label{tab:compression-error}
    \vspace{-0.2cm}
\end{table}

\subp{Compression error.} We measure the compression error in vNMSE~\cite{vargaftik2021drive}. We calculate the error via recording the output activation after each attention layer with and without compression, using the equation $\mathbb{E}(||o_l-\hat{o_l}||^2 / ||\hat{o_l}||^2)$ over all tokens in different samples and averaged over all layers.
%$l$ which is $o_l=\text{softmax}(QK^T/\sqrt{d})V$ with and without compression (denoted as $o_l$ and $\hat{o_l}$ respectively), and the vNMSE is calculated as $\mathbb{E}(||o_l-\hat{o_l}||^2 / ||\hat{o_l}||^2)$ over tokens in different samples and is averaged over all layers.
As shown in~\Cref{tab:compression-error}, \sysname and its variants outperform standard INT8, INT4, and CacheGen quantization schemes by at least one order of magnitude respectively. This reflects why \sysname's prototypes achieve less accuracy loss than other compressed baseline approaches accordingly.

\begin{figure*}
    \centering
    \vspace{0.1cm}
    \begin{minipage}[t]{0.34\linewidth}{
		\vspace{-0.00in}
		\begin{center}
		\includegraphics[width=\textwidth, ]{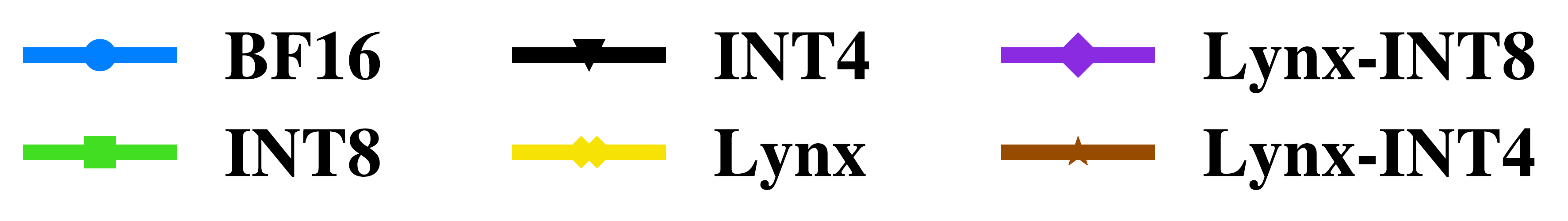}
		\end{center}
		}
        \end{minipage}

    % \hspace{-0.2cm}
        \subfigure[MMLU LLaMA]{
		\begin{minipage}[t]{0.25\linewidth}{
		\vspace{-0.00in}
		\begin{center}
		\includegraphics[width=\textwidth, ]{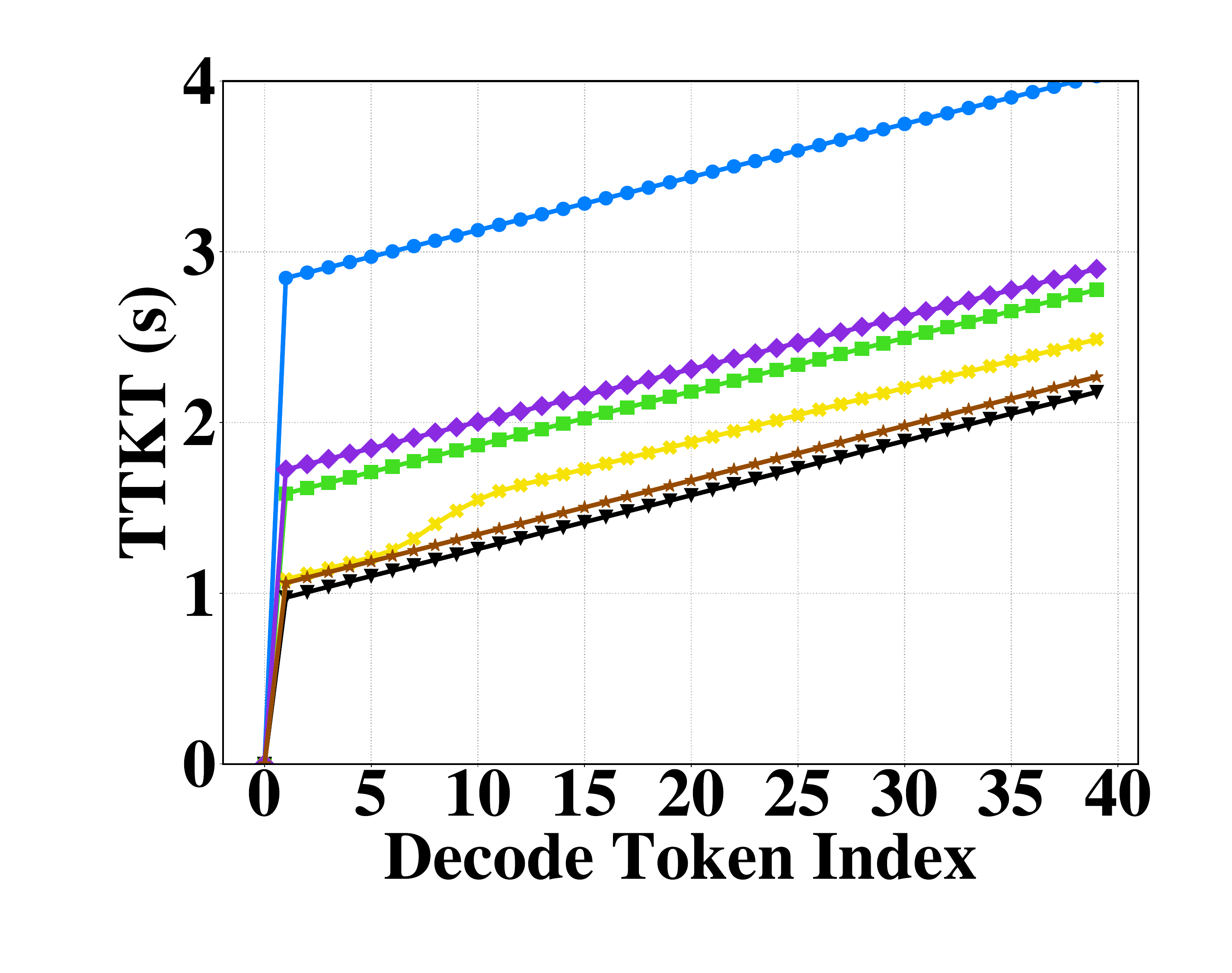}
		\end{center}
            \vspace{-0.1cm}
		}
		\label{subfig:ttkt-mmlu-llama}
		\end{minipage}
	}
    \hspace{-0.2cm}
        \subfigure[Needle LLaMA]{
		\begin{minipage}[t]{0.25\linewidth}{
		\vspace{-0.00in}
		\begin{center}
		\includegraphics[width=\textwidth, ]{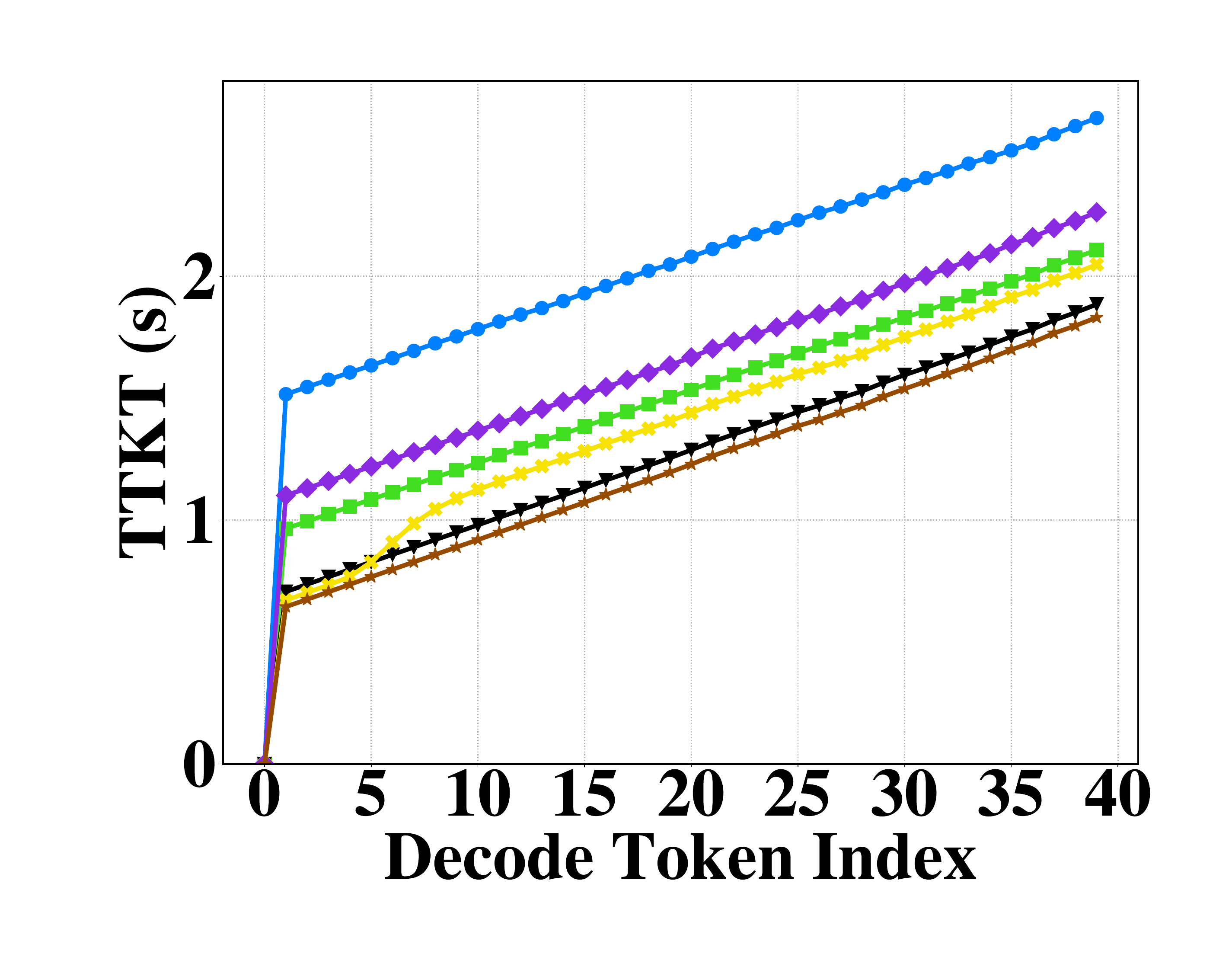}
		\end{center}
        \vspace{-0.1cm}
		}
		\label{subfig:ttkt-needle-llama}
		\end{minipage}
	}
    \hspace{-0.2cm}
        \subfigure[QMSum Mistral]{
		\begin{minipage}[t]{0.25\linewidth}{
		\vspace{-0.00in}
		\begin{center}
		\includegraphics[width=\textwidth, ]{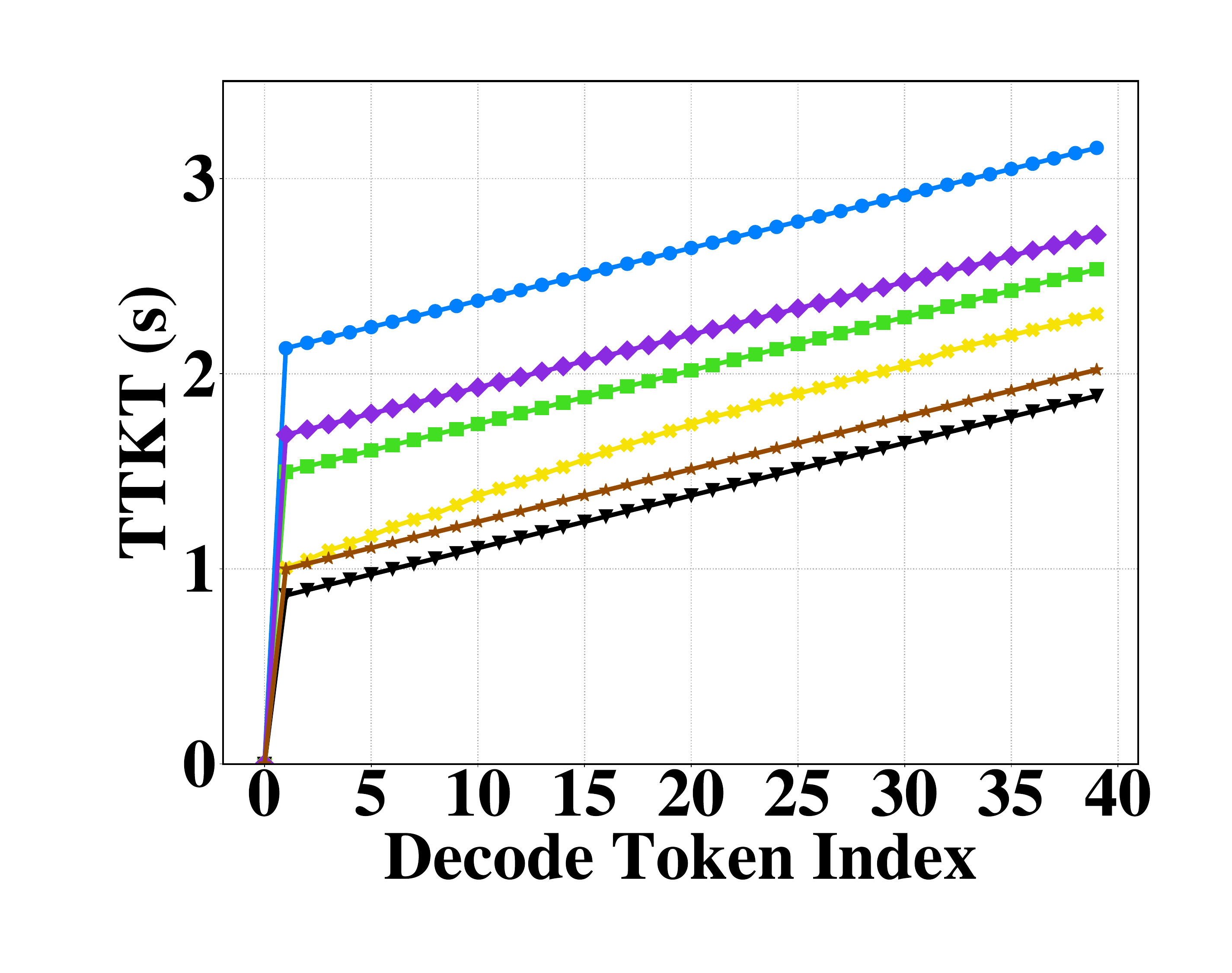}
		\end{center}
        \vspace{-0.1cm}
		}
		\label{subfig:ttkt-mistral-qmsum}
		\end{minipage}
	}

    \centering
    \hspace{-0.2cm}
        \subfigure[MMLU Qwen]{
		\begin{minipage}[t]{0.25\linewidth}{
		\vspace{-0.00in}
		\begin{center}
		\includegraphics[width=\textwidth, ]{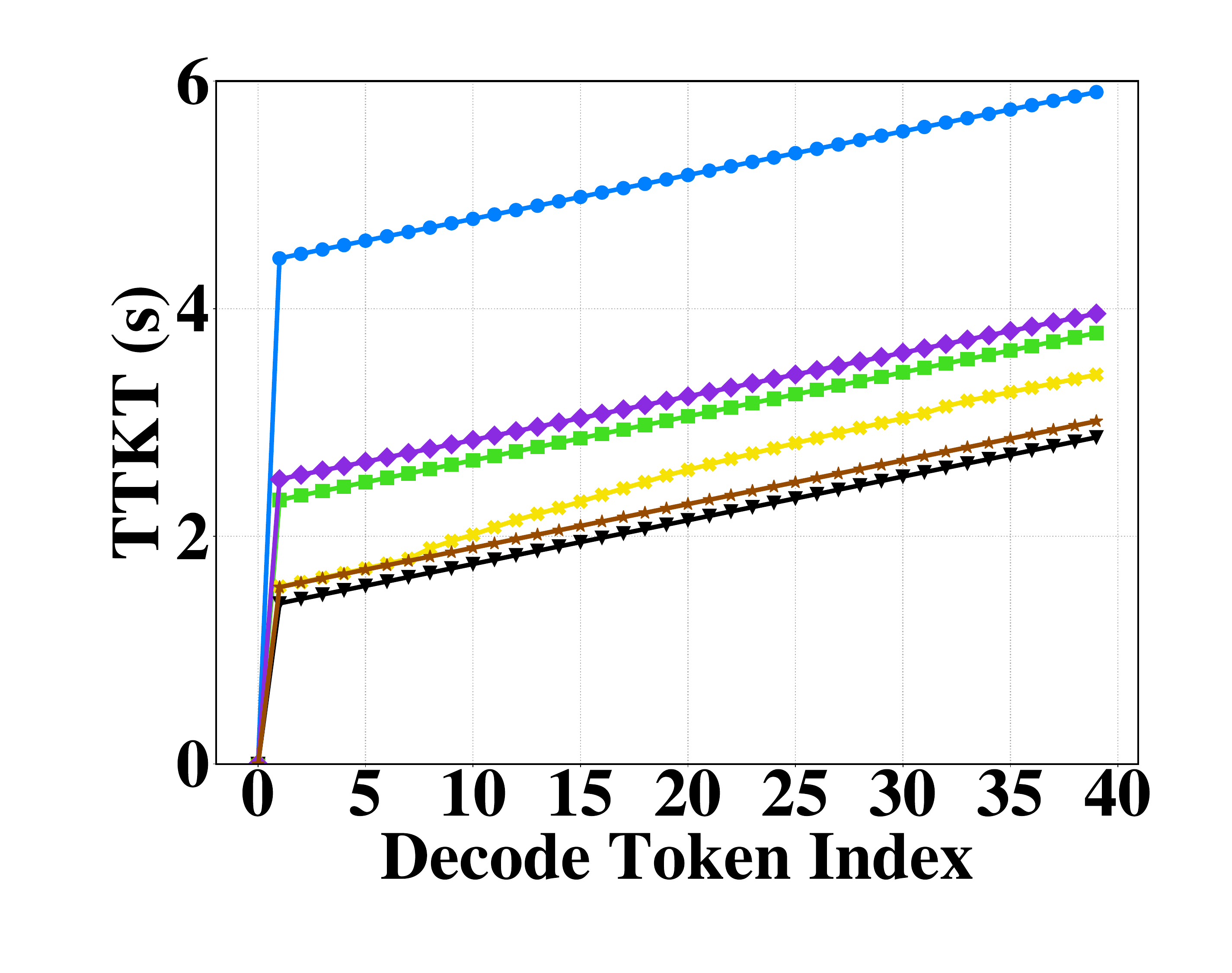}
		\end{center}
        \vspace{-0.1cm}
		}
		\label{subfig:ttkt-mmlu-qwen}
		\end{minipage}
	}
    \hspace{-0.2cm}
        \subfigure[Needle Mistral]{
		\begin{minipage}[t]{0.25\linewidth}{
		\vspace{-0.00in}
		\begin{center}
		\includegraphics[width=\textwidth, ]{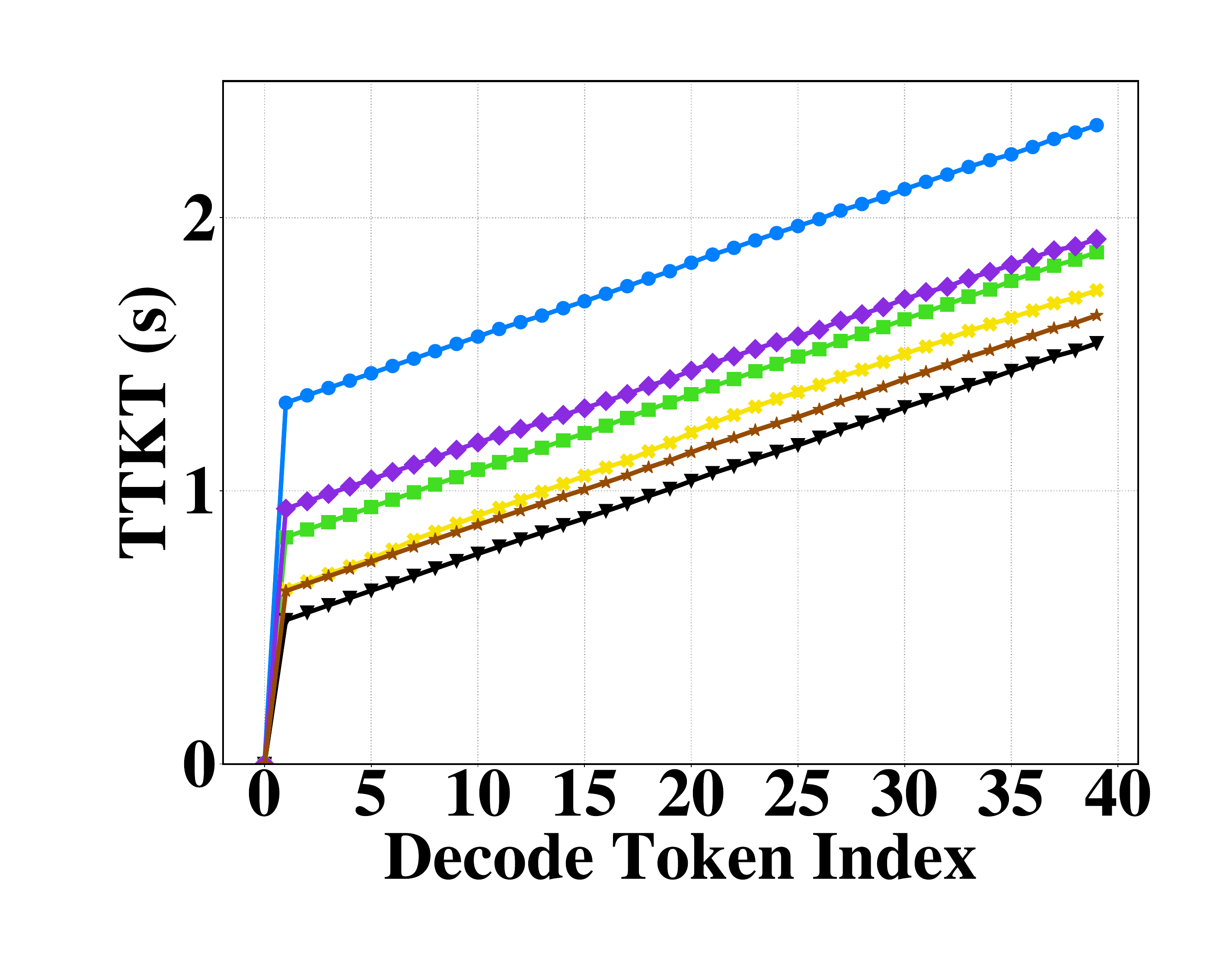}
		\end{center}
        \vspace{-0.1cm}
		}
		\label{subfig:ttkt-mistral-needle}
		\end{minipage}
	}
    \hspace{-0.2cm}
        \subfigure[QMSum Qwen]{
		\begin{minipage}[t]{0.25\linewidth}{
		\vspace{-0.00in}
		\begin{center}
		\includegraphics[width=\textwidth, ]{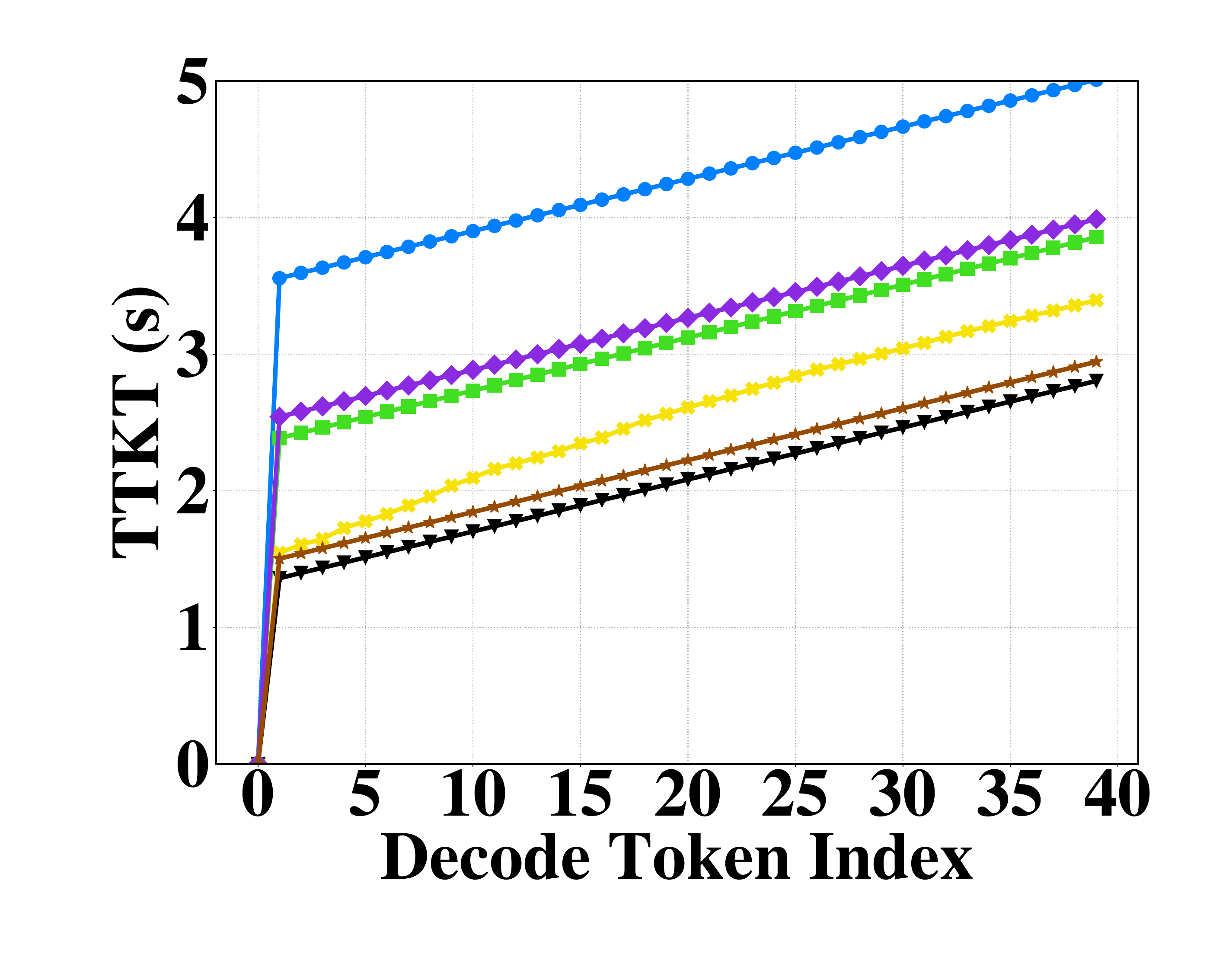}
		\end{center}
        \vspace{-0.1cm}
		}
		\label{subfig:ttkt-qmsum-qwen}
		\end{minipage}
	}

    \vspace{-0.2cm}
    \caption{End-to-end inference latency in terms of Time-to-Kth-token (TTKT) across different workloads. \sysname incurs only marginal overhead compared to its standard quantization counterparts.}
    \label{fig:ttkt-e2e}
    \vspace{-0.0cm}
\end{figure*}

\subsection{End-to-end inference latency}\label{subsec:e2e-latency}

Next, we present the end-to-end inference latency of \sysname's prototypes, comparing the results with uncompressed BF16 and the simple quantization schemes of INT8 and INT4.

\subp{TTKT.} Figure~\ref{fig:ttkt-e2e} illustrates the time each approach requires to generate the $k$th output token. We note that TTKT grows linearly with respect to $k$ for all curves (with the same model-intrinsic tangent characterizing the model's generation speed), except for \sysname, which is subject to the speculative KV transfer and decoding process.

The key observation is that \sysname, despite transmitting both $Q_{anc}$ and $Q_{res}$ of 8-bit-per-coordinate in total over the network, achieves consistently lower TTKT than INT8. In particular, the TT1T (also known as TTFT) of \sysname is even on par with INT4, which only transmits only 4-bit-per-coordinate. For example, under the MMLU+QWen workload, \sysname achieves $0.87$s faster TT1T and $0.39$s faster TT32T than INT8, which translates to approximately $29$ and $13$ TPOT (time-per-output-token) respectively. Such a reduction is attributed to the computation-communication overlap between the speculative decoding of \sysname with 4-bit anchor stream precision and the parallel network transmission of the refined residual stream. This renders \sysname to outperform INT8 and \sysname-INT8, since the total communication time for KV transfer is the same, but \sysname leverages the waiting time to generate speculative tokens ahead of INT8 and \sysname-INT8.

\sysname's advanced quantization and speculative transfer implementations inevitably introduce extra computational overhead compared with the simple INT8 and INT4 KV quantization schemes. However, such overhead is typically of low level, \eg, $0.13$s in MMLU Qwen, compared with \sysname's improved inference latency overall.

\begin{figure}[h!]
    \centering
    \vspace{0.38in}
    \hspace{-0.2cm}
        \subfigure[Theoretical acceptance rate]{
		\begin{minipage}[t]{0.49\linewidth}{
		\vspace{-0.00in}
		\begin{center}
		\includegraphics[width=\textwidth, ]{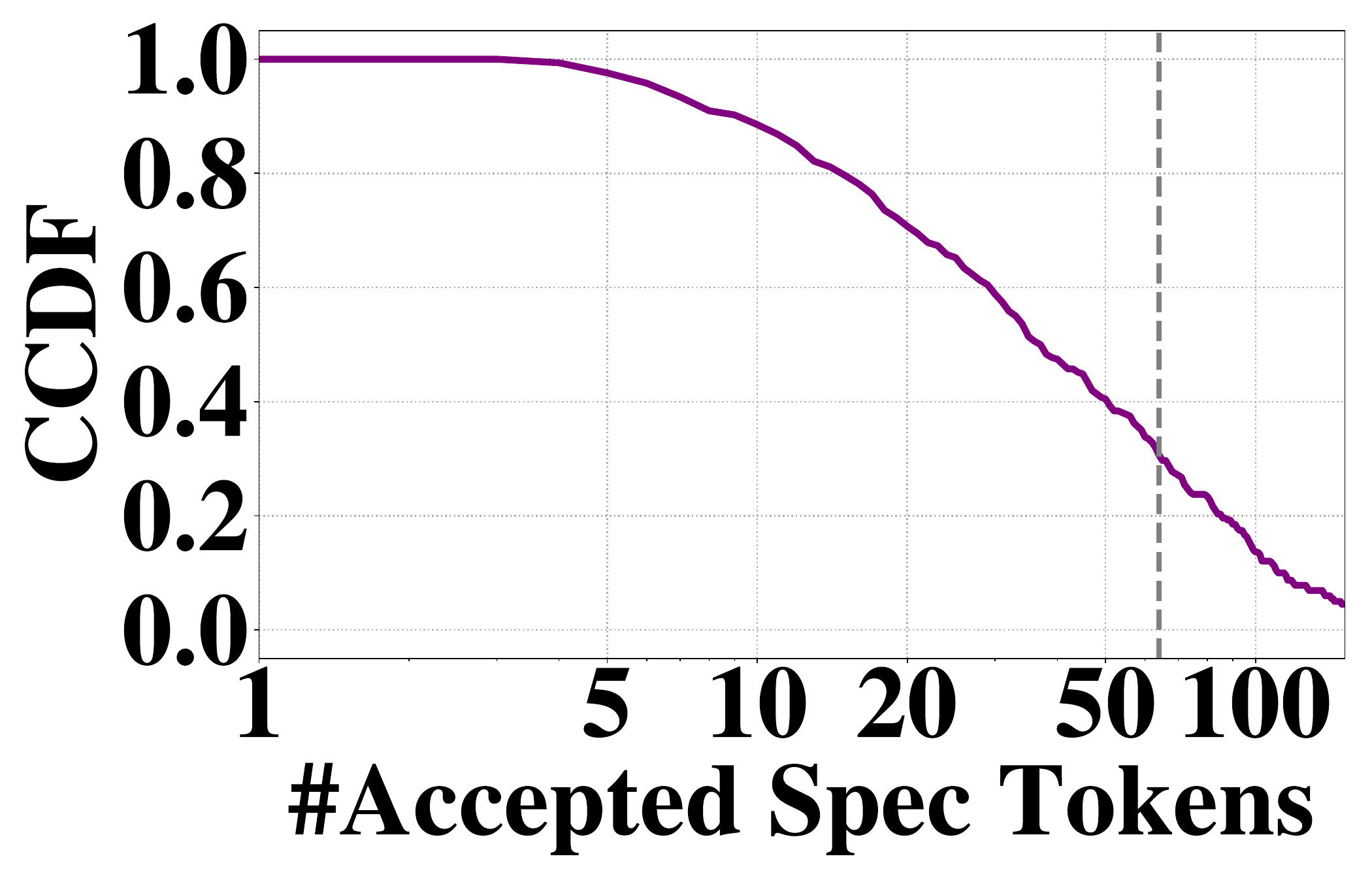}
		\end{center}
            \vspace{-0.1cm}
		}
		\label{subfig:spec-theoretical-accept-rate}
		\end{minipage}
	}
    \hspace{-0.1cm}
        \subfigure[\mbox{Actual spec-accept probability}]{
		\begin{minipage}[t]{0.46\linewidth}{
		\vspace{-0.38in}
		\begin{center}
		\includegraphics[width=\textwidth, ]{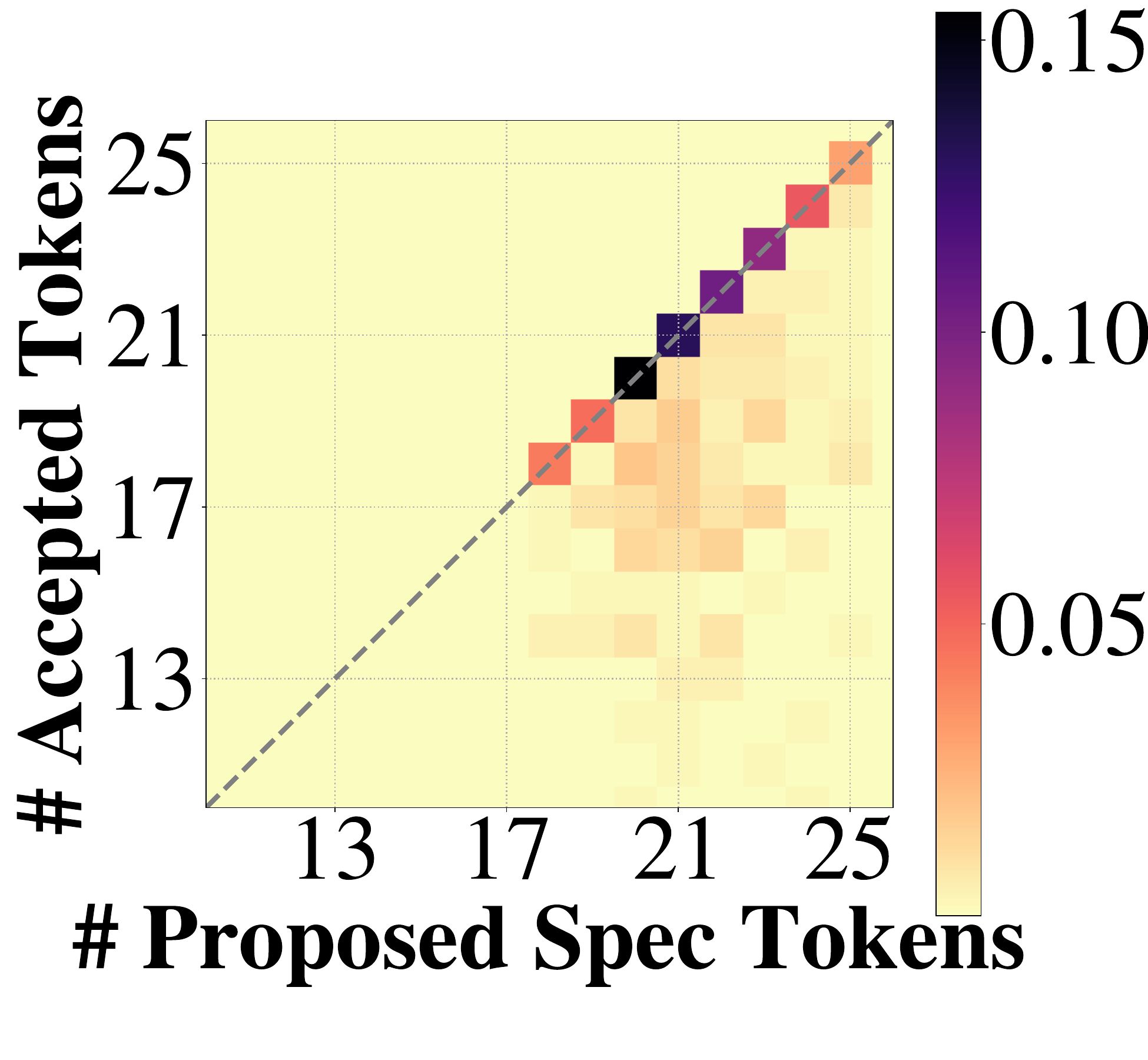}
		\end{center}
        \vspace{-0.1cm}
		}
		\label{subfig:spec-actual-accept-rate}
		\end{minipage}
	}

    \vspace{-0.2cm}
    \caption{Evaluation of \sysname's speculative tokens acceptance rate over the MMLU Qwen workload. Figure (a) depicts %the theoretical acceptance rate of speculative tokens measured as CCDF, i.e., 
    $P(\#\text{\textit{accepted}}) \ge k$, given a sufficiently large number of proposed speculative tokens. %In the analysis, the model generates a sufficient number of tokens with $\mathcal{C}_4$ (INT4 KV cache) and then verifies with $\mathcal{C}_8$ (INT8 KV cache). 
    Figure (b) depicts the heatmap showing the distribution between \#proposed speculative tokens and \#accepted tokens, where the sum of the diagonal indices equals to $64.8\%$.}
    %the probabilty that all proposed tokens are accepted (sum of diagonal entries) equals $64.8\%$.
    %Figure (b) depicts the heatmap, measuring the proportion of samples where the length of proposed speculative tokens is $x$ and the actual count of accepted tokens is $y$, measured during end-to-end experiments.}
    \label{fig:spec-acceptance}
    \vspace{0.2cm}
\end{figure}

\subp{The acceptance rate of the speculative tokens.} As mentioned earlier, \sysname's improved TTKT performance compared to INT8 KV transfer schemes stems from the number of \textit{accepted speculative tokens} generated in parallel with the residual stream's transmission. 
Theoretically, the number of speculative tokens is upper-bounded by the constant time of the residual stream's transmission divided by the TPOT of the model. Therefore, it is crucial for \sysname to ensure high \textit{acceptance rate} of the speculative tokens.

%
% 21.427698574338084 20.340122199592667
% 
Figure~\ref{fig:spec-acceptance} depicts that \sysname achieves a sufficiently high acceptance rate in our end-to-end experiments. In the MMLU Qwen workload, we find that the model generates on average $21.43$ speculative tokens, of which $19.38$ tokens are accepted; with $64.8\%$ probability, the whole sequence of the speculative tokens is fully accepted. This acceptance rate is sufficient to fully hide the residual KV transfer latency in all evaluated settings, explaining why \sysname consistently outperforms INT8 despite transmitting the same total number of bits. We further evaluate the theoretical acceptance rate by measuring the length of the accepted prefix when generating a sufficiently large number of speculative tokens using the anchor stream KV cache. As shows in Figure~\ref{subfig:spec-theoretical-accept-rate}, there is an $88\%$ probability that at least $10$ tokens are accepted, and a $70\%$ probability that at least $20$ tokens are accepted.

%This shows that even if a whole speculative sequence is not fully accepted, usually only a few last tokens are discarded and the rest of the prefix tokens are accepted. We also study the theoretical acceptance rate where the model generates sufficiently many speculative tokens and we calculate the number of tokens in its prefix that are accepted. As shown in Figure~\ref{subfig:spec-theoretical-accept-rate}, there is a high probability of $88\%$ that at least $10$ tokens are accepted, and for $20$ tokens the probability remains as high as $70\%$. These high acceptance rate results lay a solid foundation on how many tokens \sysname's speculative KV transfer can generate ahead of time than INT8 and \sysname-INT8.

% \paragraph{Takeaways.} 

\subsection{Scalability analysis}\label{subsec:exp-scalability}

\begin{figure}[h!]
    \centering
    \begin{minipage}[t]{0.7\linewidth}{
		\vspace{-0.00in}
		\begin{center}
		\includegraphics[width=\textwidth, ]{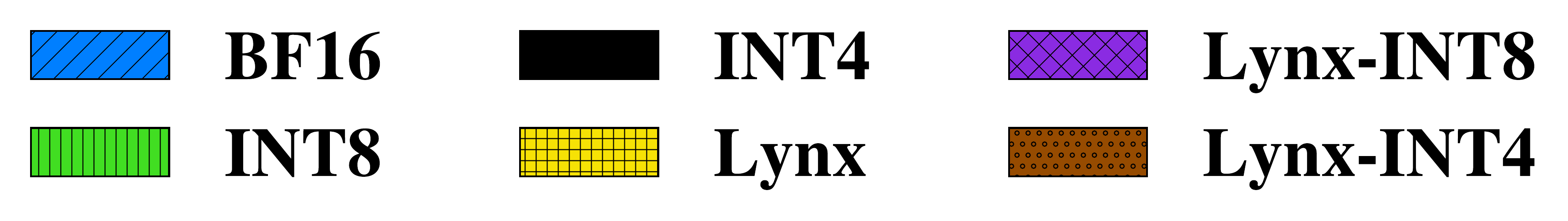}
		\end{center}
		}
        \end{minipage}

    \hspace{-0.2cm}
        \subfigure[Varying context length]{
		\begin{minipage}[t]{0.49\linewidth}{
		\vspace{-0.00in}
		\begin{center}
		\includegraphics[width=\textwidth, ]{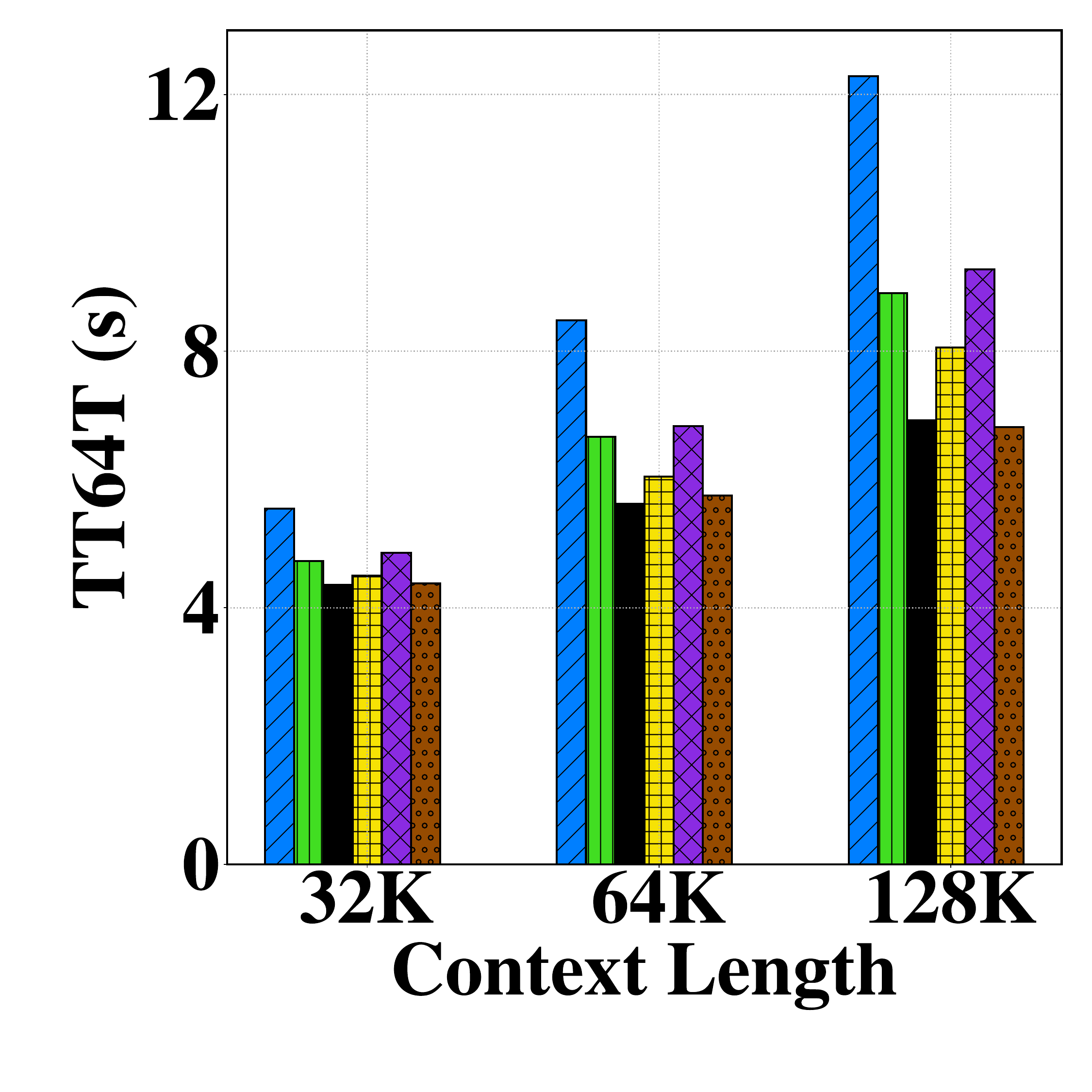}
		\end{center}
            \vspace{-0.1cm}
		}
		\label{subfig:ablation-ttkt-length}
		\end{minipage}
	}
    \hspace{-0.23cm}
        \subfigure[Varying bandwidth]{
		\begin{minipage}[t]{0.49\linewidth}{
		\vspace{-0.00in}
		\begin{center}
		\includegraphics[width=\textwidth, ]{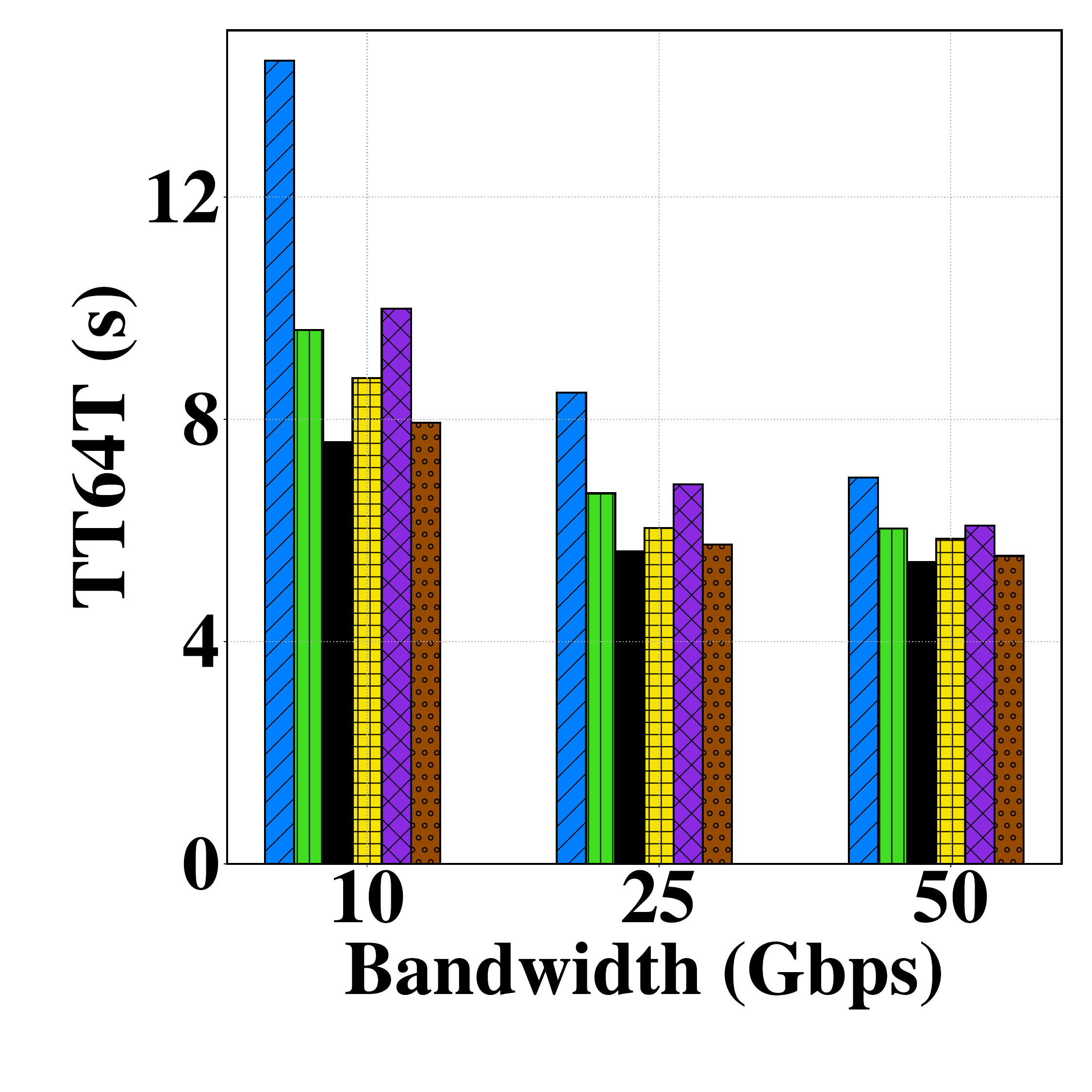}
		\end{center}
        \vspace{-0.1cm}
		}
		\label{subfig:ablation-ttkt-bandwidth}
		\end{minipage}
	}

    \vspace{-0.1cm}
    \caption{TT64T for the MMLU LLaMA workload across varying context length and bandwidth. \sysname incurs only marginal overhead compared to its standard quantization counterparts.}
    \label{fig:ablation-ttkt}
    % \vspace{0.1cm}
\end{figure}

\begin{figure}[h!]
    \centering
    \begin{minipage}[t]{0.9\linewidth}{
		\vspace{-0.00in}
		\begin{center}
		\includegraphics[width=\textwidth, ]{exp_figures/mmlu_pro_llama_accuracy_legend.pdf}
		\end{center}
		}
        \end{minipage}

    \hspace{-0.2cm}
        \subfigure[32K tokens]{
		\begin{minipage}[t]{0.32\linewidth}{
		\vspace{-0.00in}
		\begin{center}
		\includegraphics[width=\textwidth, ]{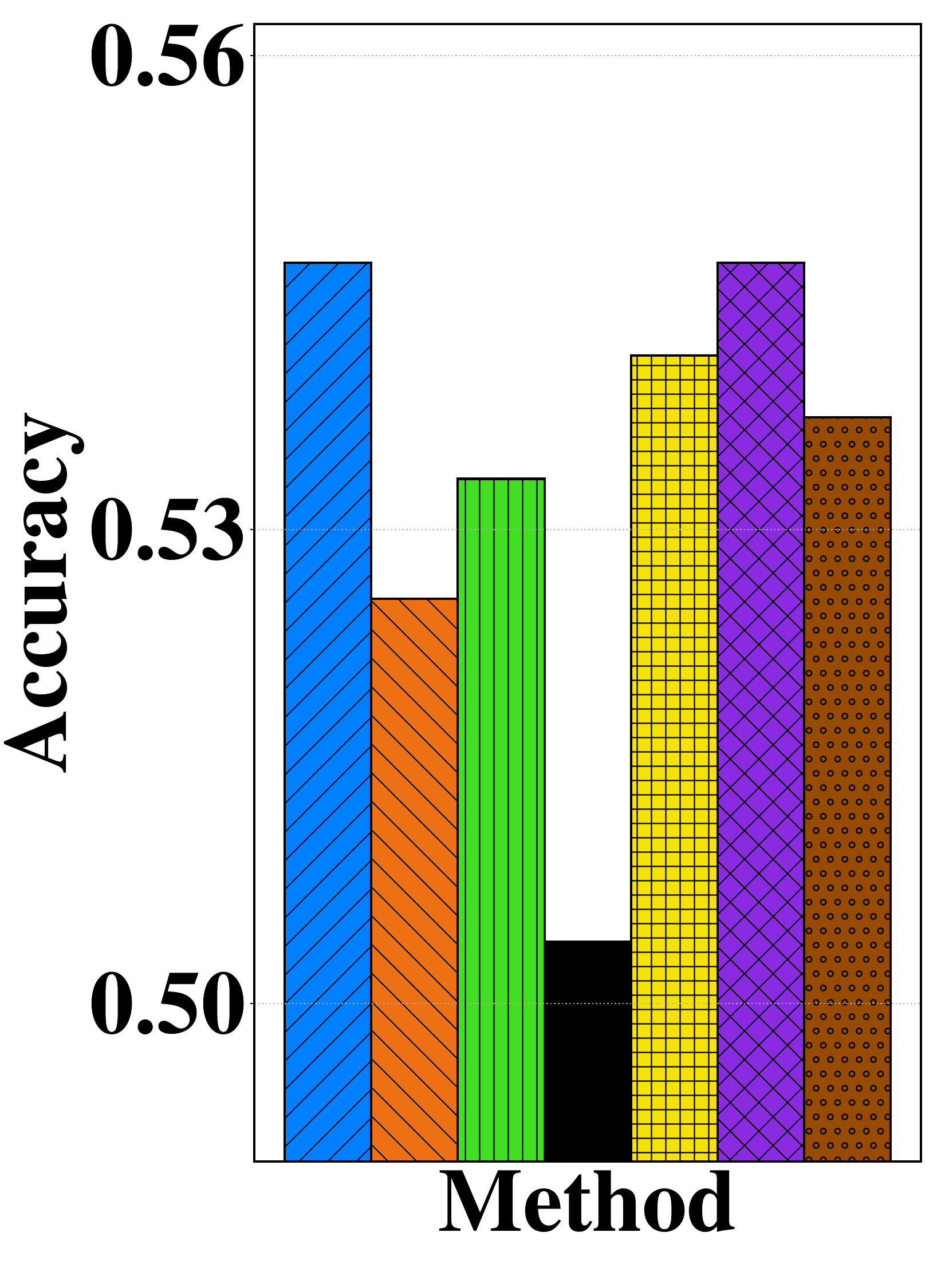}
		\end{center}
            \vspace{-0.1cm}
		}
		\label{subfig:accuracy-mmlu-llama-25-32k}
		\end{minipage}
	}
    \hspace{-0.23cm}
        \subfigure[64K tokens]{
		\begin{minipage}[t]{0.325\linewidth}{
		\vspace{-0.00in}
		\begin{center}
		\includegraphics[width=\textwidth, ]{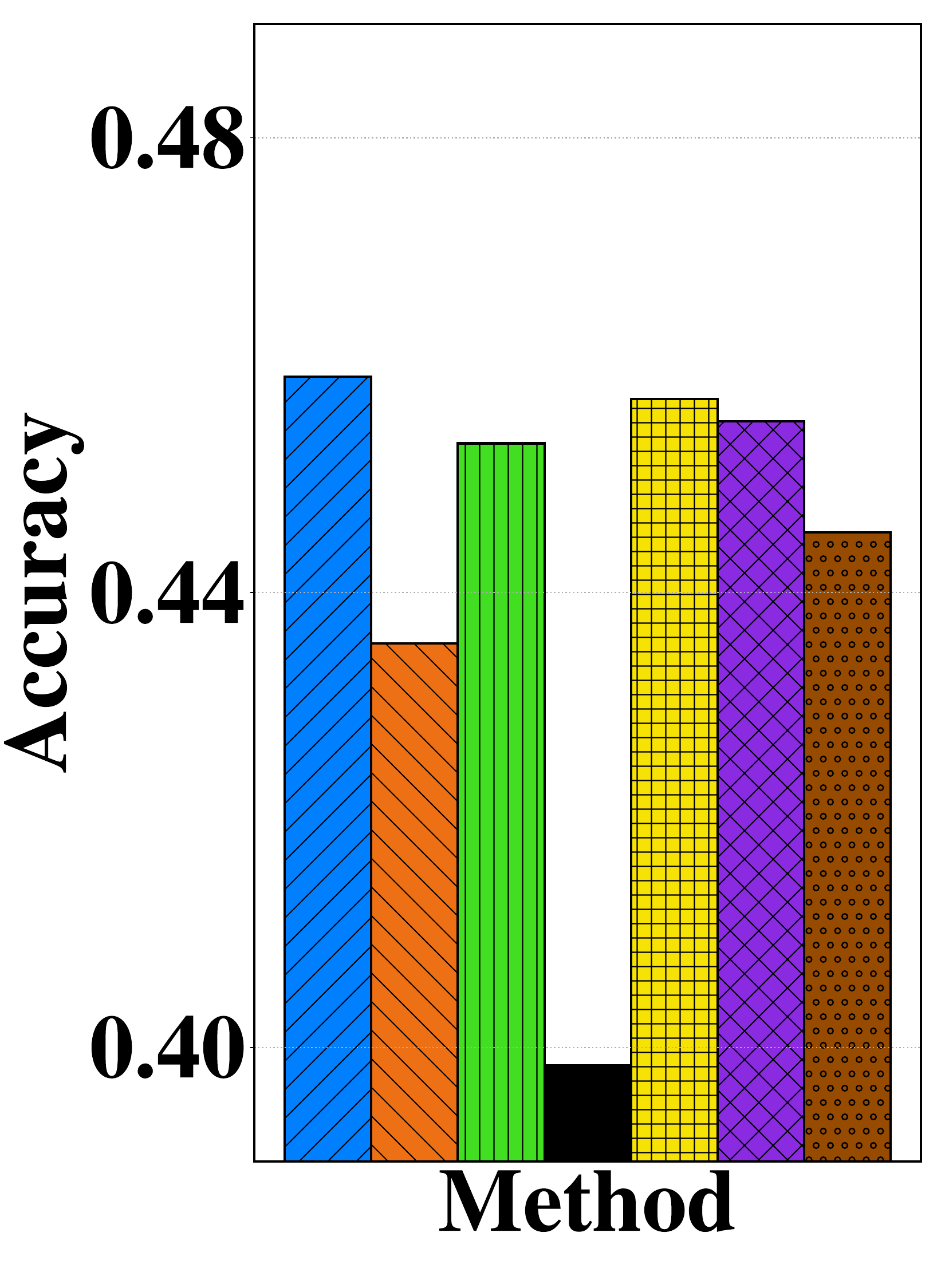}
		\end{center}
        \vspace{-0.1cm}
		}
		\label{subfig:accuracy-mmlu-llama-64k-25}
		\end{minipage}
	}
    \hspace{-0.23cm}
        \subfigure[128K tokens]{
		\begin{minipage}[t]{0.32\linewidth}{
		\vspace{-0.00in}
		\begin{center}
		\includegraphics[width=\textwidth, ]{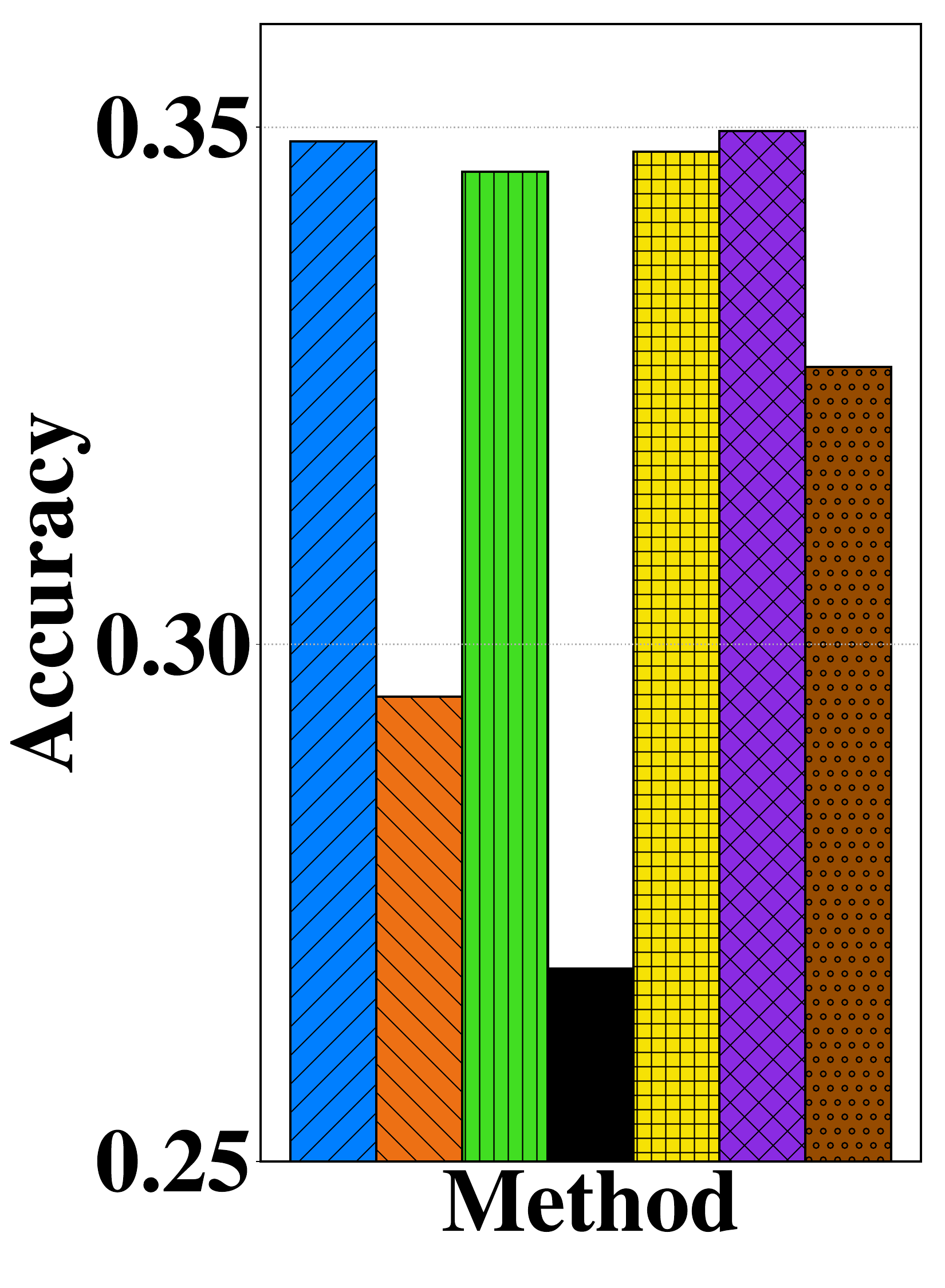}
		\end{center}
        \vspace{-0.1cm}
		}
		\label{subfig:accuracy-mmlu-llama-128k-25}
		\end{minipage}
	}

    \vspace{-0.3cm}
    \caption{The inference accuracy varying the context length from 32K tokens to $128$K tokens, running LLaMA 8B over the MMLU dataset.}
    \label{fig:ablation-accuracy-length}
    % \vspace{0.1cm}
\end{figure}

\subp{Scaling context lengths to $128$K tokens.} We evaluate \sysname's scalability against long context by varying the context length of our MMLU-Pro workload from $32K$ tokens to $128$K tokens, and plot the TTKT results (where $K=64$) and inference accuracy, in ~\Cref{fig:ablation-ttkt} and ~\Cref{fig:ablation-accuracy-length} respectively. Here the bandwidth is set as $25$Gbps.

Regarding inference accuracy, \sysname and \sysname-INT8's accuracy remain consistently within the $\pm 0.5\%$ range of BF16's accuracy. Conversely, both CacheGen and INT4 suffer from increasingly severe accuracy degradation with respect to longer contexts, \eg, from $2.5\%$ to $5.3\%$ for CacheGen raising the context length from $32$K tokens to $128$K tokens. The vanilla INT8 scheme also begins to experience an accuracy loss of up to $0.9\%$, due to its compression algorithm being less accurate than \sysname's optimized algorithm (Table~\ref{tab:compression-error}).

Regarding inference latency, \sysname's TT64T outperforms vanilla INT8 consistently across different context lengths, and such gain increases from $0.22$s with $32$K tokens to $0.84$s with $128$K tokens. In general, \sysname yields more inference latency reduction for longer contexts, as the residual stream's transmission lasts longer and \sysname thus generates more speculative tokens in parallel. We note that the end-to-end results have already taken into account both the computational overhead of decompression and time for correcting the rejected speculative tokens.

\subp{Scaling bandwidth to $50$Gbps.} We also evaluate how bandwidth affects inference latency of \sysname and baseline schemes. We choose to experiment on the MMLU LLaMA workload with $64$K context length, and the results are depicted in Figure~\ref{subfig:ablation-ttkt-bandwidth}. It can be observed that \sysname consistently achieves lower TT64T than INT8 across the range of $10$ to $50$Gbps bandwidth, though arguably such gain shrinks from $0.86$s to $0.18$s for $10$ and $50$Gbps respectively. In general, \sysname's improved inference latency is proportional to the saved KV transfer overhead of the residual stream compared with vanilla INT8, and \sysname yields maximum speedups under the scenario of low bandwidth and long context.

\section{Related work}
%\gingfung{maybe we don't need this anymore? since this is discussed in background}
%\subp{Disaggregated prefill-decode inference.} Disaggregated prefill–decode (PD) inference~\cite{zhong2024distserve} separates the compute-heavy prefill stage from the latency-sensitive auto-regressive decode stage into different nodes to improve throughput and resource utilization. One key challenge with disaggregated inference is the transfer of key–value (KV) caches from prefill workers to decode workers, which studies have shown can dominate end-to-end latency~\cite{li2025flowkv, chen2024kvdirect}. Recent prevailence of long-context workloads~\cite{yao2025cacheblend} further intensify the bottleneck, as the KV cache size grows accordingly~\cite{kwon2023efficient}. 

\subp{KV Compression in Transfer.} Prior works propose to accelerate KV transfer by compressing the KV cache data to a more compacted form. One major class of compression schemes is KV quantization~\cite{xiao2023smoothquant, lin2025qserve, liu2024kivi, hooper2024kvquant, zhang2025hack}, which converts high-precision representations (such as FP16 and BF16) into low-precision formats such as INT8, INT4, and microscaling floating-point formats~\cite{opencompute}. 

An alternative to quantization is sparsification-based compression. ScissorHands~\cite{liu2023scissorhands} and SparQ~\cite{ribar2024sparq} retrieve the KV cache corresponds to only the most important tokens to be transmitted. The above two categories are lossy compression schemes, in that they introduce compression error, which in turn results in inaccurate inference and degraded inference quality. In general, lossy compression cannot achieve the best of both worlds: high compression ratio and high inference quality. Conversely, \sysname achieves both by overlapping KV transfer with the computation of decoding, hiding partially the KV transmission latency of the less important bits, and by enabling decoding to proceed on a partial KV state rather than waiting for KV completeness. 

In parallel with lossy compression algorithms are lossless encoding algorithms. For example, Cachegen~\cite{liu2024cachegen} proposes, atop their quantization scheme, a frequency-based arithmetic encoding scheme to re-encode the KV data into a compact bit-stream. However, such encoding is not only computationally expensive but also requires that the whole encoded data be received by the decoder before the data can be successfully, so that overlapping KV transfer with decoding becomes infeasible. Existing KV transfer techniques fundamentally trade accuracy for size reduction or compression efficiency; \sysname avoids both by overlapping transfer with speculative execution. Unlike prior speculative decoding work, which accelerates computation by approximating model execution, \sysname applies speculative execution at the communication boundary, overlapping KV transfer with decoding.

\subp{Low-precision LLM inference.} Another line of work is to use lower-precision weights and arithmetics for faster and more memory-efficient LLM inference. Weight-only quantization such as GPT-Q~\cite{frantar2022gptq} and AWQ~\cite{awq} are \textit{offline} \mbox{approaches} that take place during the post-training phase, where only the model's weight matrices are quantized to reduce the memory footprint. The objective here is to minimize the inference quality degradation caused by quantization, and the algorithm is allowed sufficiently long time for such offline optimization. Activation quantization during serving~\cite{10.1145/3676641.3716252, bondarenko2023quantizable} is of greater challenge, where the quantization takes place at the serving runtime. Such quantization requires not only accuracy in quantization but also efficiency in the low-precision Gemm computation, where hardware-friendly quantization schemes~\cite{Dubtsov2023cuBLAS12, opencompute} gain an advantage.

\section{Conclusion}

This paper presents \sysname, a KV-cache compression and transfer system that rethinks KV movement as a divisible, agile, and pipelined operation, rather than a blocking bottleneck. By combining hierarchical non-linear quantization with Anchor–Residual prioritization and speculative decoding, Lynx overlaps KV transfer with token generation, minimizing exposed communication overhead while preserving full-precision output quality.

More broadly, \sysname reframes KV transfer from a blocking prerequisite into a progressively usable resource, allowing decoding to proceed before KV completeness. Experiments demonstrates Lynx achieves INT4-level time-to-first-token while matching BF16/INT8 accuracy for up to 128K context length, significantly outperforming existing approaches. %Overall, Lynx bridges the gap between communication efficiency and accuracy-preserving KV transfer, achieving the best of both worlds. 

\sysname’s prototype uses INT4 for both the anchor and residual streams, yielding an effective INT8 representation. This design is motivated by two considerations. First, INT8 KV caches already match BF16 accuracy with negligible degradation. Second, using power-of-two bit-widths enable efficient memory-coalesced kernel implementation on modern hardware accelerators. In principle, \sysname can support alternative bit-width configurations for the anchor and residual streams (e.g., INT6 or INT2), to trade-offs inference quality and TTKT latency. We leave a systematic exploration of these configurations to future work. %By proposing that decoding need not wait for KV completness, \sysname opens a new direction for communication-aware inference system design.

Finally, while \sysname is implemented on the Ascend NPU platform, its design is hardware-agnostic and readily generalizable to other AI accelerators such as NVIDIA's GPUs. We leave such extensions to future work. By proposing that decoding need not wait for KV completeness, \sysname opens a new direction for communication‑aware inference system design. In essence, we remove a network‑induced serialization barrier by making application execution progressive with respect to communication.
% remove AI ?
% \Wenchen{Discussion that we can not only adapt to INT4/INT8 but also INT6/INT12 etc.}

\iffalse
\Wenchen{Discussion for batched request. We note that in vllm's schedule, all requests are treated fairly, which means that facing bandwidth contention between concurrent requests, the bandwidth is fair-shared, i.e., all requests will need to wait for KV cache from all workers to be transmitted before decoding. For \sysname, we 1) transmit the MSB of KV cache for all requests, 2) decoding overlapped with transmitting LSB for all request.}
\fi

\section{Acknowledgment}
\gingfung{https://conferences.sigcomm.org/sigcomm/2026/cfp/ <-- use of generative AI ? should we say it ?} \Wenchen{Space-permitted, we can add this, but this is not necessary.}
We acknowledge the use of Generative AI to refine the structure and clarity of the text. All technical analysis, experimental design, and final writing decisions
remain the sole responsibility of the authors. 
This work does not raise any ethical issues.

\newpage
\bibliographystyle{ACM-Reference-Format}
% \begin{small}
\bibliography{sigcomm26}
% \end{small}

% \newpage

% \input{appendix}

\end{document}